# The Exact Stabilization Time of Self-Stabilizing Coloring: Mean, Variance, and Distribution

**Élie de Panafieu**
Nokia Bell Labs, Paris-Saclay, France

**Fabien Mathieu**
Swapcard, Paris, France
Sorbonne Université, CNRS, LIP6, F-75005 Paris, France

**Maria Potop-Butucaru**
Sorbonne Université, CNRS, LIP6, F-75005 Paris, France

**Sébastien Tixeuil**
Sorbonne Université, CNRS, LIP6, F-75005 Paris, France
Institut Universitaire de France, Paris, France

## Abstract

The worst-case stabilization time of a self-stabilizing algorithm usually says little about its typical behavior beyond the order of magnitude. This paper revisits the classical self-stabilizing $(\Delta+1)$-coloring algorithm with vertices activated at random, one at a time (the random central daemon). Considering the number of color changes as a random variable, we study its mean, variance and, on several families of graphs, its exact distribution. The analysis rests on a reduction: a vertex changes color, at most once, exactly when it is not a local maximum of a uniformly random labeling of the vertices, maxima being taken among the neighbors that share its initial color (the monochrome initial configuration is the slowest case). What can be predicted depends on what is known of the graph. The degree distribution (the view at distance one) gives the mean. The variance needs the view at distance two, in particular the triangles, and the full law needs the whole graph. We give closed-form results, from monochrome and from uniformly random initial configurations, on paths, cycles, random trees, Erdős-Rényi graphs and bipartite random graphs. From a monochrome start, the mean stays within a factor two of the worst case, whereas the variance ranges from zero (disjoint cliques) to order $n^2$ (the star on $n$ vertices).

**Supplementary Material** Software (Source Code): https://github.com/balouf/tinto/tree/v0.1.0 archived at `swh:1:rev:79d72a3026ff486616ef0ba83b70e192de339cee`

**Acknowledgements** We thank Stephan Wagner for bibliographic references on random trees.

## 1 Introduction

A self-stabilizing algorithm reaches a legitimate configuration (for a given specification) from any initial one and remains legitimate afterwards [Alt+19, Dij74, Dol00]. At each step a *daemon* chooses which of the enabled processes move. The *stabilization time* is the number of moves to legitimacy, in the worst case over the initial configuration and the daemon's choices.

We consider the classical self-stabilizing $(\Delta+1)$-coloring rule on a graph $G=(V,E)$ whose vertices are the processes, each holding a color from a palette of at least $\Delta+1$ colors, $\Delta$ being the maximum degree [Alt+19, Dol00, GT00, HJS03]. A process whose color is shared by a neighbor is enabled and, when activated, recolors itself with any color absent from its neighborhood (such a color always exists). Under the central daemon of Dijkstra's original paper [Dij74] (exactly one enabled process per step), the worst case is easy: a recolored process is never in conflict again, so every process moves at most once [HJS03]. For any $G$, the longest execution, over initial configurations and schedules, makes $n - \kappa(G)$ moves, where $n := |V|$ and $\kappa(G)$ is the number of connected components of $G$. In other words, the worst case sees nothing of $G$ beyond $\kappa(G)$. We wonder what the graph determines once the daemon's choices are random. Beyond the worst case, self-stabilization mostly studies the expected stabilization time of randomized algorithms, Herman's token ring being the classic example [Her90]: the expectation is over their coins, under a synchronous or worst-case daemon. Here, in contrast, we analyze a deterministic algorithm under a random daemon.

To study typical behavior, we consider the random version of the central daemon that draws the next mover uniformly among the enabled processes. The stabilization time of a deterministic algorithm then becomes a random variable [DTY15, Yam+12]. Observe that the random central daemon is a scheduling abstraction, not a centralized algorithm: each process reads only its neighbors' colors, and every step is one recoloring by a process in conflict. Counted in recolorings, the stabilization time has the same law when each process is activated by the ticks of its own rate-1 Poisson clock (Proposition 1). Our objective is the law of this random variable: its mean and variance on every graph and from every initial configuration, its full distribution on several families of graphs.

The key to our approach is as follows: draw a uniformly random ordering of the processes, and let the daemon always serve the enabled process that comes first in it. The execution has the same law as under the random central daemon, and the ordering alone decides which processes move. We call a process a *peak* when it comes after every neighbor that shares its initial color. A process recolors if and only if it is not a peak, so the stabilization time is exactly $n$ minus the number of peaks of a random ordering, whatever free color each move picks.

From a monochrome start (one common color), the peaks are the local maxima of a random labeling of the graph, an object probability and combinatorics know well: on a path they are the peaks of a random permutation [DB62, Eli16], on a general graph their number is a standard example for Stein's method [BRS89], and its expectation is the Caro-Wei bound on the independence number [Car79, Wei81]. We recall the classical results where they are used (Section 3, Section 5). To our knowledge, the link between peaks and stabilization time is new, and all our contributions follow from it.

The mean alone says little: from a monochrome start, on a graph without isolated vertices, it lies between $n/2$ and $n - 1$, always within a factor two of the worst case (Proposition 3). We therefore study the whole distribution, and the more we know of the graph, the finer the prediction. The mean needs only the degree distribution (Theorem 1). The variance needs only, for each pair of processes at distance at most two, the sizes of their common and private neighborhoods, in particular the triangles (Theorem 3, Eq. (6)). The law needs the whole graph. The approach is general: the same proofs cover every graph and every initial configuration, whether monochrome, uniformly random with any number $k$ of colors, or fixed, through the subgraph of initially conflicting edges (Proposition 2); the monochrome start is the slowest for every ordering of the processes.

On graphs of bounded degree, the stabilization time is concentrated within order $\sqrt{n}$ of its mean, and is asymptotically normal as soon as its variance grows linearly (Corollary 2). For a monochrome start, the extremes are explicit (Section 4): the variance vanishes exactly on disjoint unions of cliques, never exceeds $n(n + 2)/12$, and is maximized by the star for every $n$. Finally, on five families of graphs, we push the analysis to the law itself whenever possible (Section 5). Besides paths and cycles, these include three random models: uniform random trees, Erdős-Rényi graphs $G(n,p)$ and their bipartite version. Random graphs are a common testbed for simulations in self-stabilization, yet they are rarely analyzed, a recent exception being [GZ23].

**Related work.** *Stabilization time of self-stabilizing coloring.* Self-stabilizing vertex coloring was first studied for feasibility on particular graph classes [GK93, SS93], then on arbitrary graphs with explicit stabilization bounds. Gradinariu and Tixeuil [GT00] give deterministic $(\Delta+1)$-coloring algorithms that stabilize in $O(n\Delta)$ moves under an unfair daemon, and a randomized one that stabilizes in $O((\Delta-1)\log n)$ expected steps of the distributed daemon, a step possibly moving many processes, the expectation being over the algorithm's coin flips, against the worst schedule. Hedetniemi, Jacobs and Srimani [HJS03] give two algorithms for the central daemon, one reaching a Grundy coloring in a number of moves linear in the number of edges, the other a $(\Delta+1)$-coloring in at most $n$ moves. Bernard, Devismes, Paroux, Potop-Butucaru and Tixeuil [Ber+10] analyze a probabilistic coloring algorithm on anonymous networks with one-way links under the unfair distributed daemon. The algorithm's expected stabilization time, again over its coins and from the worst initial configuration and schedule, goes from $O(n\Delta)$ moves with $\Delta+1$ colors down to $O(n)$ as the palette grows. Other work on the problem is worst-case too: unidirectional coloring [Ber+09], transformers from local or synchronous to self-stabilizing algorithms [Joh+24, LSW09], dynamic and locally iterative coloring [BEG22, FYZ25]. Two notions of expected stabilization time coexist. In the randomized algorithms above the expectation is over the algorithm's coins, under a fixed or adversarial daemon [DIM95], whereas in our work the algorithm is deterministic and the expectation is over a random daemon, possibly also over a random initial configuration or a random graph, so that for a given algorithm it is at most the worst case.

*Average-case analyses in self-stabilization.* The most studied case is Herman's randomized token circulation on rings [Her90], whose expected stabilization time under synchrony has been bounded with increasing precision [FMP04, FZ15, Kie+11, Nak05]. Fribourg, Messika and Picaronny [FMP04, FMP06] use coupling to bound two measures of randomized self-stabilizing algorithms: the expected stabilization time and the time after which stabilization has occurred with probability at least $1-\varepsilon$. Both bounds are worst case over the initial configuration, with the probability taken over the algorithm's coins. Modeling the algorithm as a Markov decision process whose policies are the daemon's choices, Fribourg and Messika [FM05] extend these bounds to arbitrary daemons, and Beauquier, Johnen and Messika [BJM06] reduce the expected stabilization time against the best or the worst daemon to a stochastic shortest-path problem. Exhaustive, symbolic or synthesis-based methods compute such recovery measures on bounded instances [ABT15, Fag+18, FB15, FBT13, Vol+22]. Turau bounds the mean time to repair a single fault for randomized coloring under synchrony [Tur18]. Instead of solving a Markov chain or decision process, or exploring bounded instances, we reduce the stabilization time to the number of peaks of a random ordering. Speculative stabilization [DG13] designs algorithms to be fast under a more likely daemon while keeping the worst-case guarantee, whereas here the algorithm is fixed and we measure how long its typical executions take. Closer to ours in spirit, Burman et al. [Bur+21] analyze leader election in population protocols under the uniform random pairwise scheduler, which is a random daemon, and Giakkoupis and Ziccardi [GZ23] give graph-dependent bounds for a synchronous randomized maximal-independent-set process from arbitrary initial states: $O(\log n)$ expected rounds on cliques and trees, polylogarithmic rounds with high probability on $G(n,p)$.

**Organization.** Section 2 fixes the model. Section 3 gives the reduction and the mean and variance theorems, Section 4 the extremal results, Section 5 treats the five families, and Section 6 discusses the other daemons and what is left open. The body proves Lemma 1, Theorem 4, Theorem 5 and the short corollaries, and sketches the other proofs. The complete proofs, the variance formulas of the $k$-colored tree and of the two random-graph families, the Poisson Galton-Watson tree, the figures and the numerical checks are in the appendix, which keeps the section order of the body. The AI-disclosure section follows the references.

## 2 Model

The rule and the terminology below are standard in self-stabilization [Alt+19, Dij74, Dol00], and self-stabilizing $(\Delta+1)$-coloring under a central daemon goes back to [GT00, HJS03]. Up to its palette, the rule below is the second algorithm of [HJS03].

**Network.** The network is an undirected graph $G = (V, E)$ with $n = |V|$ vertices. Two vertices $u, v$ are *neighbors* if $uv \in E$. The neighborhood of $u$ is $\mathcal{N}(u)$, its degree $d_u = |\mathcal{N}(u)|$, and $\Delta = \max_u d_u$ is the maximum degree. Each vertex hosts a process, and we identify the two.

**State, guard, action.** Each process holds a *color* drawn from a palette of size $q$. We consider the classical self-stabilizing coloring rule with the *conflict* (or *clash*) guard: a process is *enabled* iff it shares its color with at least one neighbor. When activated, an enabled process recolors itself with any color absent from its neighborhood. We take the palette large enough that a free color always exists, $q \geq \Delta + 1$, so an enabled process always has a color to move to. A configuration is *legitimate* iff no process is enabled, that is, iff the coloring is proper.

**Daemon.** We use the *central* (sequential) daemon, under which exactly one enabled process executes its action at each step, in its *random* version: the process is drawn uniformly at random among the enabled ones. The stabilization time $T$ is the number of activations, which is also the number of steps, until the configuration is legitimate. We write $T_n$ when the dependence on $n$ matters. By Lemma 1 below, $T$ does not depend on which free color a process picks. The random central daemon is our model, but two scheduler-free settings give the same number of recolorings in law (Proposition 1): a daemon that draws at each step a process uniformly among all $n$, enabled or not, and a network in which every process wakes up at the ticks of its own Poisson clock of rate 1 and moves if it is enabled, reading its neighbors and writing its color in one atomic step (composite atomicity).

**Rounds.** Rounds carry no information here: the set of enabled processes only shrinks along an execution (Lemma 1), so every execution from a non-legitimate configuration takes exactly one round (a round ends once every process enabled at its start has moved or ceased to be enabled), under any central daemon.

**Initial configuration and colors.** The algorithm may start from any configuration. Two reference distributions give explicit averaged formulas: the *monochrome* configuration, where all processes start with the same color, and the *random* configuration, where each process starts with an independent uniform color among $k$ colors (not necessarily a proper coloring). Here $k \leq q$, and $k = 1$ is the monochrome case. The palette size $q \geq \Delta + 1$ guarantees a free color and nothing else, so $T$ depends on $q$ only through this inequality, while $k$ enters our formulas through the probability $1/k$ that a given edge is initially in

conflict.[1]

**Variants.** The choice of guard matters, the choice of action does not. Under the Grundy-type guard, where a process is enabled as soon as its color is not the smallest one absent from its neighborhood, a process may move several times and our analysis does not apply. Under the other daemons the law of $T$ is not covered: the order of the moves is no longer uniform (locally central), or neighbors may move together (synchronous, distributed). Section 6 discusses these cases.

## 3 General theorems for the mean and the variance

Throughout, the vertices carry an initial coloring (fixed, monochrome, or i.i.d. uniform among $k$ colors) and a uniform random labeling by distinct integers, which the next lemma links to the daemon. Equivalently, the labels are i.i.d. uniform on $[0,1]$, since only their relative order matters, and the proofs use this version. A vertex $v$ is a *peak* if its label is larger than the label of every neighbor sharing its color (in the ordering of the introduction, a large label is a late turn). A vertex with no same-colored neighbor is a peak. The number of peaks is written $S$, $S_n$, $S(\sigma)$ or $S(H)$ according to what is fixed (the size, the labeling, the graph).

### 3.1 The reduction lemma

▶ **Lemma 1** (Reduction). *Fix an initial coloring, a palette $q \geq \Delta + 1$ and a free-color recoloring policy that depends only on the current configuration and on internal coins independent of the labeling. Let $\sigma$ be a labeling of $V$ by distinct integers and consider the* smallest-label schedule*, which at each step activates the enabled vertex of smallest label.*

1. *Each vertex is activated at most once, and a vertex that is not enabled never becomes enabled.*
2. *A vertex is activated if and only if it is not a peak of the initial coloring under $\sigma$. The schedule therefore performs $n - S(\sigma)$ moves, $S(\sigma)$ being the number of peaks.*
3. *If $\sigma$ is a uniform random labeling, the smallest-label schedule has the law of the random central daemon, so that $T_n = n - S_n$ in law, $S_n$ being the number of peaks of a uniform random labeling.*

**Proof.** (1) A move recolors its process with a color absent from the neighborhood, so it never creates a conflict: a process that has moved has no conflict and never regains one, and a process without conflict never acquires one. Writing $A_t$ for the set of processes activated before step $t$ and $E_t$ for the set of enabled ones, $E_t$ consists of the processes outside $A_t$ that have a same-colored neighbor outside $A_t$, and $E_{t+1} \subseteq E_t \setminus \{u_t\}$, where $u_t$ is the process activated at step $t$.

(2) If $v$ is a peak, every same-colored neighbor $w$ of $v$ has a smaller label and, as long as $v$ holds its initial color, is enabled until it acts. Under the smallest-label schedule $w$ acts before $v$ and recolors away from $v$'s color, so once all of them have acted $v$ has no conflict and is never activated. If $v$ is not a peak, it has a same-colored neighbor $u$ of larger label. The two stay in conflict until one of them acts, $u$ cannot act first, and the schedule stops after at most $n$ moves by (1), so $v$ acts.

[1] On a random graph $\Delta$ is random, but $q = n$ always works, and the peak statistic itself needs no palette constraint (Appendix E).

(3) The history of the schedule before step $t$, that is, the activated processes and the colors they chose, determines $E_t$. Given the policy's coins, independent of $\sigma$, this history is the event that $\sigma(u_s) < \sigma(x)$ for all $x \in E_s \setminus \{u_s\}$ and all $s < t$. Since $E_t \subseteq E_s \setminus \{u_s\}$ for every $s < t$ by (1), permuting the labels of $E_t$ among themselves leaves this event unchanged. Conditionally on the history, the relative order of the labels of $E_t$ is therefore uniform, and the process of smallest label in $E_t$, the one the schedule activates, is uniform on $E_t$. This is one step of the random central daemon, and the two executions have the same law by induction on $t$. ◀

The policy entered the proof only through "picks a free color", and the daemon only through the uniform order of first activations, which two scheduler-free models share.

▶ **Proposition 1** (Scheduler-free readings). *The number of moves has the law $n - S_n$ of Lemma 1 in each of the two following models. (1) At each step a process is drawn uniformly among all $n$ processes and moves if it is enabled, doing nothing otherwise. The number of steps until legitimacy is then the time needed to draw $J$ distinct processes, $J$ being the rank of the last non-peak in the order of first draws ($J = 0$ if the initial configuration is legitimate). Its mean given $J$ is $n(H_n - H_{n-J})$, with $H_m$ the $m$-th harmonic number, and since $J \le n - 1$ the mean number of steps is at most $n(H_n - 1)$, with equality on the monochrome clique. (2) Every process carries an independent Poisson clock of rate 1 and moves at each of its ticks at which it is enabled. The time until legitimacy is the first tick of the last non-peak, in the order of first ticks.*

The proof (Appendix A) labels every process by the rank of its first draw or first tick, a uniform random permutation, and observes that a process moves at its first activation iff it is not a peak, and never again.

▶ **Proposition 2** (Fixed initial configuration). *Let $\gamma$ be an initial configuration and $G_\gamma$ its* conflict graph*, the spanning subgraph of $G$ whose edges join two vertices with the same color under $\gamma$, and let $S(H)$ be the number of local maxima of a labeling on $H$, $\alpha(H)$ its independence number and $\kappa(H)$ its number of connected components.*

1. *$T(G,\gamma) = n - S(G_\gamma)$ in law: every monochrome result of this paper applies to $\gamma$ once $G$ is replaced by $G_\gamma$, and in particular $\mathbb{E}\, T(G,\gamma) = \sum_v d_v(\gamma)/(d_v(\gamma)+1)$ with $d_v(\gamma)$ the number of neighbors of $v$ sharing its color (Theorem 1 below).*
2. *$S(G_\gamma) \ge S(G)$ for every labeling, so under a common labeling the monochrome configuration stabilizes last: it has the largest $\mathbb{P}(T \ge t)$ for every $t$.*
3. *Over all central and locally central daemons (the latter activate at each step a set of pairwise non-adjacent enabled processes), random or not, the number of moves from $\gamma$ takes exactly the values $n - \alpha(G_\gamma), \ldots, n - \kappa(G_\gamma)$. The worst case over configurations and schedules is $n - \kappa(G)$.*

Parts (1) and (2) are Lemma 1 read on $G_\gamma$ and the fact that a local maximum of $G$ is one of its spanning subgraph $G_\gamma$. The comparison in (2) is in law, not in variance: on $K_{2,3}$, recoloring one vertex of the smaller side raises the variance from 9/20 to 11/16. For (3), labeling the processes of any execution in activation order makes it a smallest-label schedule by Lemma 1(1). A step of a locally central daemon can be serialized without changing the moves, and the sets of local maxima of labelings are exactly the independent sets that meet every component (Appendix A).

With $m$ the number of initially conflicting processes, the worst case reads $n - \kappa(G_\gamma) = m - \kappa'(G_\gamma)$, $\kappa'$ counting the components of $G_\gamma$ that have an edge. The folklore bound of $m - 1$ moves (in each conflict component at least one process never moves) is thus tight exactly when there is one such component. The best schedule needs $n - \alpha(G_\gamma)$ moves (NP-hard to compute) and the worst $n - \kappa(G_\gamma)$ (linear time). By part (1) the random daemon exceeds the best schedule on average by exactly $\alpha(G_\gamma) - \sum_v 1/(d_v(\gamma) + 1)$, the slack of the Caro-Wei bound.

The peaks of a monochrome labeling are also a known object elsewhere: the vertices selected in the first round of the random-priority algorithm for maximal independent set [ABI86, Lub86, Mét+11]. The same random order drives the greedy algorithms whose parallel depth is studied by [BFS12] and [FN20], where every vertex eventually acts and the depth is the quantity of interest, whereas here the quantity is the number of moves.

## 3.2 The mean

▶ **Theorem 1** (Mean). *Let $G$ be a graph on $n$ vertices, possibly random, and let $p_d$ be the probability that a uniformly random vertex has degree $d$ (averaged over $G$ when $G$ is random). Give every vertex an independent uniform color among $k$. The expected number of peaks is*

$$\mathbb{E}(S_n) = nk \sum_d \frac{p_d}{d+1} \left(1 - \left(1 - \frac{1}{k}\right)^{d+1}\right), \tag{1}$$

*and hence the mean stabilization time is $\mathbb{E}(T_n) = n - \mathbb{E}(S_n)$. For $k = 1$ this gives the monochrome case $\mathbb{E}(T_n) = n \sum_d \frac{d}{d+1} p_d$.*

The proof (Appendix A) conditions on the degree $d$ of a uniform vertex and on its number $m \sim \text{Binomial}(d, 1/k)$ of same-colored neighbors, which it beats with probability $1/(m + 1) = \int_0^1 t^m \,\mathrm{d}t$. The mean depends on the graph *only* through its degree distribution $\boldsymbol{p}$. The factor $\frac{1}{k}$ is the probability that a given edge is initially in conflict: for the mean, a $k$-coloring on $G$ behaves like the monochrome case on the random subgraph where each edge is kept with probability $\frac{1}{k}$ (on a forest the kept edges are independent and the equality holds in law, see Section 5.2). We write $\rho_k(d) = k \int_{1-1/k}^1 u^d \,\mathrm{d}u = k/(d+1)\left(1 - (1 - 1/k)^{d+1}\right)$ for the peak probability of a degree-$d$ vertex, so that $\rho_1(d) = 1/(d + 1)$.

## 3.3 The variance

By Lemma 1, $\mathrm{Var}(T_n) = \mathrm{Var}(S_n)$, and $\mathrm{Var}(S_n) = \mathbb{E}(S_n(S_n - 1)) + \mathbb{E}(S_n) - \mathbb{E}(S_n)^2$, so the variance follows from the *second factorial moment* $\mathbb{E}(S_n(S_n - 1)) = \sum_{u \neq v} \mathbb{P}(u, v \text{ both peaks})$. Whether two vertices are both peaks depends on how their neighborhoods overlap, so the variance, unlike the mean, sees more of the graph than degrees. For an ordered pair of distinct vertices $(u, v)$, let $a$ be the number of neighbors of $u$ that are neither $v$ nor a neighbor of $v$, $c$ the symmetric count for $v$, and $b$ the number of common neighbors. In the monochrome case only non-adjacent pairs contribute, since two adjacent vertices of the same color are never both peaks.

▶ **Theorem 2** (Second factorial moment, monochrome). *In the monochrome case, the two vertices of an ordered non-adjacent pair with parameters $(a, b, c)$ are both peaks with probability*

$$J(a, b, c) = \frac{1}{a + b + c + 2} \left(\frac{1}{a + b + 1} + \frac{1}{b + c + 1}\right), \tag{2}$$

*and those of an adjacent pair never are. With $p_{a,b,c}$ the probability that a random ordered pair is non-adjacent with parameters $(a, b, c)$,*

$$\mathbb{E}(S_n(S_n - 1)) = n(n-1) \sum_{a,b,c} \frac{p_{a,b,c}}{a + b + c + 2} \left(\frac{1}{a + b + 1} + \frac{1}{b + c + 1}\right). \tag{3}$$

The proof (Appendix A) conditions on the uniform labels $s, t$ of $u, v$: a private neighbor of $u$ ranks below $u$ with probability $s$, a common neighbor below both with probability $\min(s,t)$, and $J(a,b,c)$ is the integral of $s^a t^c \min(s,t)^b$ over the unit square. With $k$ colors the same computation applies to $s = (k-1+x)/k$, the probability that a given neighbor of a vertex of label $x$ fails to beat it (wrong color, or same color and lower label). Then $s$ is uniform on $[1-1/k, 1]$. A common neighbor beats neither of two same-colored vertices with probability $\min(s,t)$, and neither of two differently colored ones with probability $s+t-1$.

▶ **Theorem 3** (Second factorial moment). *With every vertex colored uniformly among $k$ colors, the two vertices of an ordered pair $u, v$ with parameters $(a,b,c)$ are both peaks with probability $\pi_{\text{same}}(a,b,c) + \pi_{\text{dist}}(a,b,c)$ if non-adjacent and $\pi_{\text{dist}}(a,b,c)$ if adjacent, where $\pi_{\text{same}}$ (resp. $\pi_{\text{dist}}$) is the probability that $u$ and $v$ are both peaks and received the same color (resp. distinct colors):*

$$\begin{aligned} \pi_{\text{same}}(a,b,c) &= k \int_{1-1/k}^{1} \int_{1-1/k}^{1} s^a t^c \min(s,t)^b \, \mathrm{d}s \, \mathrm{d}t, \\ \pi_{\text{dist}}(a,b,c) &= (k-1)k \int_{1-1/k}^{1} \int_{1-1/k}^{1} s^a t^c (s+t-1)^b \, \mathrm{d}s \, \mathrm{d}t. \end{aligned} \tag{4}$$

*Then, with $q_{a,b,c}$ the probability that a random ordered pair is adjacent with parameters $(a,b,c)$,*

$$\mathbb{E}(S_n(S_n-1)) = n(n-1) \sum_{a,b,c} \left[ p_{a,b,c}(\pi_{\text{same}} + \pi_{\text{dist}}) + q_{a,b,c}\pi_{\text{dist}} \right]. \tag{5}$$

*At $k=1$ one has $\pi_{\text{dist}} = 0$ and $\pi_{\text{same}} = J$, recovering Theorem 2. Both integrals are elementary, and their values as finite sums are given in Appendix A, where the theorem is proved.*

▶ **Corollary 1** (Pairs without common neighbor). *At $b=0$ both integrands factorize, with $\rho_k(a) = k\int_{1-1/k}^{1} s^a \, \mathrm{d}s$ as in Theorem 1: $\pi_{\text{same}}(a,0,c) = \rho_k(a)\rho_k(c)/k$ and $\pi_{\text{dist}}(a,0,c) = (k-1)\rho_k(a)\rho_k(c)/k$. Two non-adjacent vertices without common neighbor are thus independent peaks, for every $k$, and the covariance of two peak indicators vanishes beyond distance two. If $G$ is triangle-free, adjacent pairs have $b=0$ and contribute $(k-1)\rho_k(a)\rho_k(c)/k$ each.*

The variance is therefore a sum over the pairs at distance at most two,

$$\mathrm{Var}(T) = \sum_v \rho_k(d_v)(1-\rho_k(d_v)) + \sum_{u \neq v,\, \mathrm{dist}(u,v) \leq 2} [\mathbb{P}(u, v \text{ both peaks}) - \rho_k(d_u)\rho_k(d_v)], \tag{6}$$

and it sees the profile $(a,b,c)$ of these pairs, in particular the triangle count, and nothing else: the cycle $C_6$ and two disjoint triangles, both 2-regular, have the same mean 4 and variances $12/45$ and 0. The maximum degree nevertheless bounds the fluctuations.

▶ **Corollary 2** (Concentration and normal approximation). *Let $G$ have maximum degree $\Delta$, with a fixed initial configuration or a uniform random $k$-coloring.*

1. *For every $t > 0$, $\mathbb{P}(|T - \mathbb{E}\, T| \geq t) \leq 2\exp(-2t^2/(n(\Delta+1)^2))$.*
2. *The Kolmogorov distance (the supremum distance between distribution functions) between the law of $(T - \mathbb{E}\, T)/\sqrt{\mathrm{Var}(T)}$ and the standard normal law is at most $C\Delta^4 n/\mathrm{Var}(T)^{3/2}$ for an absolute constant $C$. Along a sequence of graphs of bounded degree, $T_n$ is therefore asymptotically normal whenever $\mathrm{Var}(T_n)/n^{2/3} \to \infty$, in particular when $\mathrm{Var}(T_n)$ is of order $n$.*

The proof (Appendix A) is the bounded differences inequality [McD89], the rank and color of a vertex influencing at most $\Delta+1$ peak indicators, and the normal approximation theorem for sums of variables with a dependency graph (a graph on the summands in which families with no edge between them are independent) [BR89, Rin94], indicators at distance at least 3 depending on disjoint sets of variables. The number of local maxima of a random labeling is a standard example of this method [BRS89, Gol05].

## 4 Extremal graphs

Fix the number of vertices $n$. We work in the monochrome regime ($k = 1$), where a peak is a *local maximum*. By Proposition 2 the statements apply to any fixed configuration, once the graph is replaced by its conflict graph restricted to the initially conflicting processes and $n$ by their number. The statements about the mean extend to every $k$, while those about the variance are specific to $k = 1$, a single edge having positive variance once $k \geq 2$. We use repeatedly that $\mathbb{E}(S) = \sum_v 1/(d_v + 1)$ (Theorem 1, the Caro-Wei bound) and that the peaks form an independent set containing the vertex of largest label, so $1 \leq S \leq \alpha(G)$. Throughout, $G$ is a *fixed* graph and the randomness is the labeling (randomized analogues in Appendix B).

### 4.1 Extremes of the mean

▶ **Proposition 3** (Mean extremes). *Among graphs on $n$ vertices with no isolated vertex, $\mathbb{E}(T)$ is minimized by the perfect matching (for even $n$, the graph of $n/2$ disjoint edges, every vertex of degree 1), at $n/2$, and maximized by the clique $K_n$, at $n - 1$. Hence $\mathbb{E}(T) \in [n/2, n - 1]$. For odd $n$ the minimum is $(3n + 1)/6$, attained by a path on three vertices together with a perfect matching of the other vertices. The extremal graphs are the same for every $k$.*

The proof (Appendix B) is the termwise bound $1/n \leq 1/(d_v + 1) \leq 1/2$, and the same for $\rho_k$, which is decreasing.

### 4.2 Zero variance

▶ **Theorem 4** (Zero variance). *In the monochrome case, $\mathrm{Var}(T) = 0$ if and only if $G$ is a disjoint union of cliques, in which case $T = n - \kappa(G)$ deterministically, $\kappa(G)$ being the number of connected components.*

**Proof.** If $G$ is a disjoint union of cliques, every labeling of a clique has a unique local maximum, so $S = \kappa(G)$ under any labeling. Conversely, a graph that is not such a union contains an induced path $abc$. Give $a, b, c$ the three largest labels and fix the others. Every other vertex has the same peak status in all such labelings, while ranking $b$ below both $a$ and $c$ gives two peaks among $\{a, b, c\}$ and ranking it above both gives one, so $S$ is not constant. ◀

### 4.3 Maximal variance: the star

At the other end the variance can be of order $n^2$. On the star the stabilization time is essentially the rank of the center's label, so it does not concentrate at all. To avoid clashing with the peak count $S$, we write $\mathrm{St}_n$ for the star on $n$ vertices (one center, $n - 1$ leaves), that is, the complete bipartite graph $K_{1,n-1}$.

▶ **Proposition 4** (Star law and variance). *On the monochrome star $\mathrm{St}_n$, $\mathbb{P}(T = j) = 1/n$ for $j \in \{1, \dots, n - 2\}$ and $\mathbb{P}(T = n - 1) = 2/n$. Hence*

$$\mathbb{E}(T) = (n-1)\frac{n+2}{2n},$$
$$\mathrm{Var}(T) = (n-1)\frac{2n^2+5n-6}{6n} - (n-1)^2\frac{(n+2)^2}{4n^2} = \frac{n^2}{12} - \frac{13}{12} + o(1). \tag{7}$$

A leaf is a peak iff its label exceeds the center's, so $T$ is the label rank of the center, except that the rank $n$ also gives $T = n-1$ (Appendix B). Since every leaf is compared with the same center, $S$ is almost uniform on $\{1, ..., n-1\}$. The matching upper bound for general graphs comes from a classification of the pairwise covariances, computed from Theorem 2 (proof in Appendix B).

▶ **Lemma 2** (Pairwise covariance). *Let $u, v$ be distinct vertices with peak indicators $X_u, X_v$. If $u$ and $v$ are adjacent, $\mathrm{Cov}(X_u, X_v) = -\mathbb{E}(X_u)\,\mathbb{E}(X_v) < 0$. If they are non-adjacent with parameters $(a, b, c)$, then*

$$\mathrm{Cov}(X_u, X_v) = \frac{b}{(a+b+1)(b+c+1)(a+b+c+2)}, \tag{8}$$

*that is $b\,\mathbb{E}(X_u)\,\mathbb{E}(X_v)$ divided by the number of vertices at distance at most one from $u$ or from $v$. Hence $\mathrm{Cov}(X_u, X_v) = 0$ when $b = 0$; $\mathrm{Cov}(X_u, X_v) = 1/(2(c+2)(c+3))$ when $(a, b) = (0, 1)$, which is $1/12$ at $c = 0$ and at most $1/24$ otherwise; $\mathrm{Cov}(X_u, X_v) \le 1/24$ when $b = 1$ and $(a, c) \ne (0, 0)$; and $\mathrm{Cov}(X_u, X_v) \le b/((b+1)^2(b+2)) \le 1/18$ when $b \ge 2$. In particular $\mathrm{Cov}(X_u, X_v) \le 1/12$ always, with equality exactly when $u$ and $v$ are two leaves hanging from the same vertex.*

▶ **Theorem 5** (Variance upper bound). *For every graph on $n$ vertices, monochrome, $\mathrm{Var}(T) \le n(n+2)/12$. Hence the maximum variance over $n$-vertex graphs is $n^2/12(1+o(1))$, with sharp leading constant $1/12$, attained to leading order by the star ($\mathrm{Var}(T) = n^2/12 - 13/12 + o(1)$ on $\mathrm{St}_n$ by Eq. (7)).*

**Proof.** Write $\mathrm{Var}(T) = \mathrm{Var}(S) = \sum_v \mathrm{Var}(X_v) + \sum_{u \ne v} \mathrm{Cov}(X_u, X_v)$. Each $\mathrm{Var}(X_v) = \frac{1}{d_v+1}\left(1 - \frac{1}{d_v+1}\right) \le 1/4$, so the first sum is at most $n/4$. In the second, adjacent ordered pairs contribute negatively and, by Lemma 2, each of the at most $n(n-1)$ ordered pairs contributes at most $1/12$, so the sum is at most $n(n-1)/12$. Adding, $\mathrm{Var}(T) \le n/4 + n(n-1)/12 = n(n+2)/12$. The star's variance Eq. (7) has the same leading term $n^2/12$, so the constant is sharp. ◀

▶ **Theorem 6** (Exact variance maximizer). *For every $n \ge 3$, among all graphs on $n$ vertices, $\mathrm{Var}(T)$ is maximized by the star $\mathrm{St}_n$ and by no other graph.*

The proof (Appendix B) sorts the vertices into leaves (degree one) and the others, bounds the variance through Lemma 2 in terms of the number of leaves and of the largest number of leaves at a single vertex, and compares the bound with the star: it falls short of the star's variance for every $n \ge 7$ unless the graph is the star itself, and exhaustive enumeration settles $n \le 6$. Isolated vertices are peaks of every labeling, so a graph that has some is compared with a smaller star.

## 5 Applications to graph families

The mean of each family comes from Theorem 1 alone and the variance from Theorem 3 through the local formula Eq. (6), and only $G(n, p)$ has triangles, that is, adjacent pairs with common neighbors. On paths, cycles and random trees the variance grows linearly in $n$ and, from a monochrome start, the law is asymptotically normal. In the two Erdős-Rényi models at fixed $p$, the expected number of peaks stays bounded and the peak count converges in law without normalization. The companion code checks the formulas below numerically (Appendix G).

## 5.1 Path and cycle

The cycle $C_n$ ($n \geq 3$) is 2-regular, and the path $P_n$ ($n \geq 2$) has two vertices of degree 1 and $n-2$ of degree 2. Both are triangle-free for $n \geq 4$. Monochrome, they are the classical settings of peaks in permutations [CS74, DB62], the simplest consecutive pattern [Eli16, EN12], and in cyclic permutations [Dom+22]. The monochrome means, variances and central limit theorems below are classical [FKL22, WS96], the $k$-colored statements are new.

▶ **Corollary 3** (Path and cycle, mean). *Under a uniform random $k$-coloring,*

$$\mathbb{E}\big(T_{C_n}\big) = n\Big(\frac{1}{k} - \frac{1}{3k^2}\Big), \qquad \mathbb{E}\big(T_{P_n}\big) = n\Big(\frac{1}{k} - \frac{1}{3k^2}\Big) - \Big(\frac{1}{k} - \frac{2}{3k^2}\Big). \tag{9}$$

*In the monochrome case ($k=1$) these are $2n/3$ and $(2n-1)/3$, and for three colors the path gives $(8n-7)/27$.*

**Proof.** By Theorem 1, with $\rho_k(2) = 1 - 1/k + 1/(3k^2)$ and $\rho_k(1) = 1 - 1/(2k)$. ◀

▶ **Corollary 4** (Path and cycle, variance). *Under a uniform random $k$-coloring, with $n \geq 5$ for the cycle and $n \geq 4$ for the path,*

$$\begin{aligned} \mathrm{Var}\big(T_{C_n}\big) &= n\Big(\frac{1}{k} - \frac{2}{k^2} + \frac{4}{3k^3} - \frac{13}{45k^4}\Big), \\ \mathrm{Var}\big(T_{P_n}\big) &= \mathrm{Var}\big(T_{C_n}\big) - \Big(\frac{1}{k} - \frac{3}{k^2} + \frac{8}{3k^3} - \frac{32}{45k^4}\Big). \end{aligned} \tag{10}$$

*In the monochrome case these are* $\mathrm{Var}\big(T_{C_n}\big) = 2n/45$ *and* $\mathrm{Var}\big(T_{P_n}\big) = (2n+2)/45$.

This follows from Eq. (6) by a census of the pairs at distance one and two (Appendix C). The two families share the leading coefficient and differ by a constant boundary term, exactly as for the mean.

Beyond the moments, the path and the cycle have explicit laws, read off generating functions as follows. We root each path at an endpoint, so that labeled monochrome paths correspond bijectively to permutations. For a family of labeled, possibly colored, graphs let $\mathcal{F}(z,v) = \sum_n (z^n/n!) \sum_{G \text{ of size } n} v^{\mathrm{peaks}(G)}$, the labels $1, \dots, n$ of a labeled graph serving as $\sigma$, and write $[z^n]\mathcal{F}$ for the coefficient of $z^n$ in $\mathcal{F}$. Then $\mathbb{E}\big(v^{S_n}\big) = [z^n]\mathcal{F}(z,v)/[z^n]\mathcal{F}(z,1)$, which gives the law of $S_n$ and of $T_n = n - S_n$.

▶ **Theorem 7** (Path and cycle, full law). *The generating function of monochrome paths (permutations by peaks, a peak being a local maximum with endpoints included) is*

$$P(z,v) = \frac{v \tan\big(z\sqrt{v-1}\big)}{\sqrt{v-1} - \tan\big(z\sqrt{v-1}\big)}, \tag{11}$$

*the solution of the Riccati equation $\partial_z P = v + 2P + P^2$, $P(0,v) = 0$ [Ent69, Zhu16]. The generating functions of $k$-colored paths and of monochrome oriented cycles are*

$$\begin{aligned} \frac{kP(z,v)}{1-(k-1)P(z,v)} \quad \text{and} \quad Y(z,v) &= z + \int_0^z P(y,v)\,\mathrm{d}y \\ &= -\log\left(\cos\big(z\sqrt{v-1}\big) - \frac{\sin\big(z\sqrt{v-1}\big)}{\sqrt{v-1}}\right), \end{aligned} \tag{12}$$

*and that of $k$-colored oriented cycles is*

$$kz + \int_0^z (1+(k-1)v)P_{\mathrm{same}}(y,v)\,\mathrm{d}y + \int_0^z (2+(k-2)v)P_{\mathrm{dist}}(y,v)\,\mathrm{d}y, \tag{13}$$

*where $P_{\mathrm{same}}$ and $P_{\mathrm{dist}}$, rational in $P$ and given in Appendix C, count $k$-colored paths whose two endpoints have the same, respectively distinct, colors.*

Cycles are oriented, which does not change the peak count, and the series describe $C_n$ for $n \geq 3$. Up to the factor $v$ due to the endpoints, $P$ is the generating function of interior peaks (OEIS A008303). $Y$ counts cyclic peaks [Eli16, Ent69]. Opening a cycle at its minimum is the idea of [ES21], whose theorem does not cover peaks. The $k$-colored series are derived in Appendix C.

▶ **Corollary 5** (Central limit theorem on the path and the cycle). *On the path and on the cycle, for every $k \geq 1$, $(T_n - \mathbb{E}(T_n))/\sqrt{\mathrm{Var}(T_n)}$ converges in distribution to a standard normal law, with the mean and variance of Corollary 3 and Corollary 4.*

This is Corollary 2 with $\Delta = 2$: the variance coefficient $(45k^3 - 90k^2 + 60k - 13)/(45k^4)$ of Corollary 4 is positive for every $k \geq 1$ (it is $2/45$ at $k = 1$ and its numerator has derivative $15(3k-2)^2$), and the Kolmogorov distance to the Gaussian is $O(1/\sqrt{n})$. For the monochrome path this is the theorem of Warren and Seneta [WS96].

## 5.2 Random trees

We take the uniform random rooted labeled tree on $n$ vertices (a *Cayley tree*, not to be confused with the infinite regular tree). There are $n^{n-1}$ such trees. A peak depends only on the labeling, not on the root, so the law of the peak count is the same for the $n^{n-2}$ unrooted labeled trees. Let $C(z) = \sum_{m\geq 1} m^{m-1} z^m/m!$, with $C = ze^C$, be the Cayley tree function [Drm09, FS09]. Marking every peak by $v$, the generating function of monochrome trees by their peak count is

$$F(z,v) = C\big(ze^{(v-1)z}\big) + (v-1)z. \tag{14}$$

▶ **Theorem 8** (Random tree, exact law). *Monochrome, the number of rooted labeled trees of size $n$ with exactly $s$ peaks is $n!/s!\left\{{n-1 \atop n-s}\right\}$, where $\left\{{a \atop b}\right\}$ is the Stirling number of the second kind, so*

$$\mathbb{P}(S_n = s) = \frac{1}{n^{n-1}}\frac{n!}{s!}\left\{{n-1 \atop n-s}\right\}. \tag{15}$$

This is Theorem 5.2.1 of [Sei11], stated there for local minima. The law is the classical law of the number of *leaves* of a uniform rooted labeled tree [FS09, Rén59], and Eq. (14) is the leaf generating function, the solution of $F = z(e^F + v - 1)$. See [Pan21] for the law of other small patterns in rooted trees.

▶ **Proposition 5** (Peaks and leaves). *On a uniform random rooted labeled tree the number of peaks and the number of* leaves *(nodes with no children) have the same distribution.*

This is Corollary 5.2.2 of [Sei11], which also gives a direct bijection. The moments follow, and the mean and the variance are explicit under a random coloring as well.

▶ **Corollary 6** (Random tree, mean and variance). *Monochrome, $\mathbb{E}(S_n) = n\big(1-\frac{1}{n}\big)^{n-1}$ and $\mathbb{E}(S_n(S_n-1)) = n(n-1)\big(1-\frac{2}{n}\big)^{n-1}$, so*

$$\begin{aligned} \mathbb{E}(T_n) &= n - n\Big(1-\frac{1}{n}\Big)^{n-1} = \Big(1-\frac{1}{e}\Big)n - \frac{1}{2e} + o(1), \\ \mathrm{Var}(T_n) &= (e^{-1} - 2e^{-2})n + O(1) \approx 0.0972\, n. \end{aligned} \tag{16}$$

*Under a uniform random $k$-coloring, since the degree of a fixed vertex of a uniform labeled tree is* $1+\mathrm{Binomial}(n-2,1/n)$,

$$\mathbb{E}(T_n) = n - nk\int_{1-1/k}^{1} t\Big(\frac{n-1+t}{n}\Big)^{n-2}\,\mathrm{d}t = \big(1-e^{-1/k}\big)n + O(1), \tag{17}$$

*and*

$$\mathrm{Var}(T_n) = \big(e^{-1/k} - (1+k^{-2})e^{-2/k}\big)\,n + O(1), \tag{18}$$

*both reducing to Eq. (16) at* $k=1$.

The monochrome moments are read off Eq. (14). The $k$-colored mean combines Theorem 1 with the degree law of a vertex of a Cayley tree, and the $k$-colored variance Theorem 3 with the joint degree law of two vertices at distance one or two. The exact value of this variance is a double integral (Proposition 10 in Appendix D).

▶ **Theorem 9** (Central limit theorem on the random tree). *Monochrome, on the uniform random labeled tree* $(T_n - \mathbb{E}(T_n))/\sqrt{\mathrm{Var}(T_n)}$ *converges in distribution to a standard normal law as* $n\to\infty$, *with* $\mathbb{E}(T_n)$ *and* $\mathrm{Var}(T_n)$ *given by Corollary 6.*

This is the classical central limit theorem for the number of leaves of a conditioned Galton-Watson tree [Drm09], imported through Proposition 5 as in [Sei11] (Corollary 5.2.3). Appendix D also treats the Poisson Galton-Watson tree, which gives the Cayley tree once conditioned on its size (Proposition 11).

The Cayley analysis extends to rooted labeled trees with an arbitrary offspring distribution, and the $k$-colored law to a functional equation (Proposition 12 in Appendix D), which Lagrange inversion solves in the Cayley case.

▶ **Theorem 10** (Random tree, exact law under a $k$-coloring). *Under a uniform random $k$-coloring of the uniform random rooted labeled tree on $n$ vertices, with $k\ge 2$ and $F$ the monochrome generating function Eq. (14),*

$$\mathbb{E}\big(v^{S_n}\big) = \frac{n!}{(k-1)k^{n-1}n^n}[z^n]e^{(k-1)nF(z,v)}. \tag{19}$$

Each non-root vertex has its parent's color with probability $1/k$, independently of the others: the conflicting edges are a random subset of the tree edges, and the $k$-colored peak count is in law the monochrome peak count of this random forest. The trees containing a given spanning forest with component sizes $c_1,\ldots,c_m$ number $n^{m-2}c_1\cdots c_m$ [Moo70, Pit99], and summing over the forests gives $e^{(k-1)nF}$ (Appendix D).

## 5.3 Erdős-Rényi $G(n,p)$

In $G(n,p)$ [JŁR00] a uniform vertex has degree $d\sim\mathrm{Binomial}(n-1,p)$. The peak count has an explicit generating function. List the vertices by decreasing label, let $h_j$ be the number of vertices preceding the $j$-th one that share its color, and attribute each edge to its endpoint of smaller label: the $j$-th vertex is a peak iff none of its $h_j$ edges to these vertices is present, and these edge sets are disjoint for distinct $j$, so that, given the colors, the $j$-th vertex is a peak independently of the others with probability $(1-p)^{h_j}$. Marking peaks by $v$, $\mathbb{E}\big(v^{S_n}\big) = \mathbb{E}\prod_{j=1}^{n}\big(1+(v-1)(1-p)^{h_j}\big)$. Monochrome, $h_j=j-1$, so $S_n$ has probability generating function

$$G_{n,p}(v) = \prod_{j=1}^{n}\big(1+(v-1)(1-p)^{j-1}\big). \tag{20}$$

With $k$ colors, $h_j \sim \mathrm{Binomial}(j-1, 1/k)$. Conditioning instead on the color composition $(n_1, ..., n_k)$, which is multinomial, gives the product of the monochrome series of the color classes,

$$G_{n,p,k}(v) = \frac{1}{k^n} \sum_{n_1+\cdots+n_k=n} \binom{n}{n_1, \cdots, n_k} \prod_{i=1}^{k} G_{n_i,p}(v). \tag{21}$$

These series, and all random-graph formulas below, describe the joint law over the graph, the coloring and the labeling (Appendix E).

▶ **Proposition 6** (Erdős-Rényi, mean). *Under a uniform random $k$-coloring and a palette $q \geq \Delta + 1$,*

$$\mathbb{E}(S) = \frac{k}{p}\left(1 - \left(1 - \frac{p}{k}\right)^n\right), \qquad \mathbb{E}(T) = n - \frac{k}{p}\left(1 - \left(1 - \frac{p}{k}\right)^n\right). \tag{22}$$

*Moreover the mean at $(n, p, k)$ equals the mean at $(n, p/k, 1)$: for the mean, though not for the variance, a random $k$-coloring at edge probability $p$ behaves like the monochrome case at edge probability $p/k$.*

The mean is $\sum_j \mathbb{E}\left[(1-p)^{h_j}\right] = \sum_{j=1}^{n} (1 - p/k)^{j-1}$. The variance is in Appendix E in two closed forms: a direct one, from the same reading of the series (Proposition 13), and one that averages Theorem 3 over the multinomial neighborhood profile of a random pair (Proposition 14), which the bipartite computation reuses. At fixed $p$ the expected number of peaks tends to $k/p$ and the variance tends to a constant, instead of growing linearly in $n$.

▶ **Proposition 7** (Erdős-Rényi, limiting law). *Monochrome, $S_n = \sum_{i=0}^{n-1} B_i$ with $B_i \sim$ $\mathrm{Bernoulli}((1-p)^i)$ independent, $B_i$ being the indicator that the vertex with exactly $i$ higher-labeled vertices is a peak. As $n \to \infty$:*

1. *For* fixed $p \in (0, 1)$, *$S_n$ converges in distribution to the non-degenerate integer-valued law of $S_\infty = \sum_{i \geq 0} B_i$, with probability generating function $\mathbb{E}\left(v^{S_\infty}\right) = \prod_{i \geq 0}(1 + (v - 1)(1-p)^i)$ and mean $1/p$, so that the stabilization time is $n$ minus an $O(1)$ random quantity that never becomes deterministic.*
2. *For $p = p_n \to 0$ with $np_n \to \infty$, $(S_n - \mathbb{E}(S_n))/\sqrt{\mathrm{Var}(S_n)}$ converges to a standard normal law, with $\mathbb{E}(S_n) \sim 1/p$ and $\mathrm{Var}(S_n) \sim 1/(2p)$.*

*With $k$ initial colors, $S_\infty$ is replaced by the sum of $k$ independent copies of itself in the first regime, and in the second the normal limit persists, with $\mathbb{E}(S_n) \sim k/p$ and $\mathrm{Var}(S_n) \sim k/(2p)$.*

The proof (Appendix E) reads the factorization of Eq. (20) as a sum of independent indicators, applies Borel-Cantelli at fixed $p$ and the central limit theorem for $p \to 0$, and conditions on the color classes for $k \geq 2$.

## 5.4 Bipartite graphs

For the bipartite Erdős-Rényi model $G(n_1, n_2, p)$, a uniform vertex lies in the first side (degree $\mathrm{Binomial}(n_2, p)$) with probability $n_1/n$ and in the second side (degree $\mathrm{Binomial}(n_1, p)$) with probability $n_2/n$, where $n = n_1 + n_2$.

▶ **Proposition 8** (Bipartite Erdős-Rényi, mean). *Under a uniform random $k$-coloring and a palette $q \geq \Delta + 1$,*

$$\mathbb{E}(T) = n - \frac{k}{p}\left(\frac{n_1}{n_2+1}\left(1 - \left(1 - \frac{p}{k}\right)^{n_2+1}\right) + \frac{n_2}{n_1+1}\left(1 - \left(1 - \frac{p}{k}\right)^{n_1+1}\right)\right). \tag{23}$$

The proofs are in Appendix F, with the variance, which follows from the same averaging over the multinomial neighborhood counts as for $G(n, p)$, split according to the sides of the pair (Proposition 15). The peak count again has a product structure, now conditional on how the two sides and the colors interleave in the label order.

▶ **Proposition 9** (Bipartite Erdős-Rényi, generating function and limit law). *List the vertices by decreasing label and let $h_j$ be the number of vertices preceding the $j$-th one that lie on the other side of the bipartition and share its color. Conditionally on the sequence of sides and colors, the $j$-th vertex is a peak independently of the others with probability $(1-p)^{h_j}$, so $\mathbb{E}\big(v^{S_n}\big) = \mathbb{E}\prod_{j=1}^{n}\big(1+(v-1)(1-p)^{h_j}\big)$, the outer expectation being over the uniform interleaving of $n_1$ and $n_2$ sides and over the colors. If $p$ is fixed and $n_1/n \to \lambda \in (0,1)$, then, monochrome, $S_n$ converges in distribution to an almost surely finite random variable $S'_\infty$, with*

$$\mathbb{E}(S'_\infty) = \frac{\lambda}{(1-\lambda)p} + \frac{1-\lambda}{\lambda p}, \tag{24}$$

*and with $k$ initial colors the limit is the sum of $k$ independent copies of $S'_\infty$.*

In the regime $p \to 0$ the law should again be Gaussian, but the interleaving contributes to the variance at the same order as the conditional fluctuations, and we leave this regime open.

## 6 Conclusion and future work

Everything here follows from Lemma 1, which holds for every initial configuration and makes the monochrome one the slowest for every labeling. We leave open the central limit theorem for the $k$-colored tree, which the linear variance Eq. (18) and the $n$-th power in Eq. (19) suggest, and the regime $p \to 0$ of the bipartite model.

On the model side, the Grundy-type guard and the daemons beyond the central one change the dynamics. Under the locally central daemon the executions, once serialized, and the range of the number of moves are the same as under the central daemon (Proposition 2), but a random locally central daemon no longer induces a uniform order, and its mean number of moves is open. Under the synchronous daemon and an adversarial distributed daemon the deterministic rule does not stabilize in anonymous networks, which is why the algorithms of [GT00] and [Ber+10] use identifiers or coins. Since it converges whenever the moves are serialized, it is weak-stabilizing, and under a randomized distributed daemon it stabilizes with probability one [DTY15]. The law of its number of moves is then open. Finally, the reduction needs only that a process which has moved never moves again and never enables a neighbor. Under the Grundy-type guard a process may move several times, and what replaces the peak count for such rules is open.

## AI disclosure

We disclose that Claude (Anthropic) was used as an assistive tool during the preparation of this manuscript. Under the authors' direction, it helped with editing and drafting parts of the text, with the companion verification code, and with checking and writing up some derivations. The authors reviewed and verified all AI-assisted output and take full responsibility for the content of this paper.

## Appendix

The appendix mirrors the structure of the paper. Appendices A and B restate the results whose proofs were deferred, keeping their numbers, and the other appendices refer to the body statements by number. All add the results that the body only mentions.

# A Proofs for Section 3

▶ **Proposition 1** (restated). *Under a daemon drawing uniformly among all processes, with null activations, and under independent Poisson clocks, the number of moves has the law* $n - S_n$.

**Proof of Proposition 1.** Label every process by the rank of its first draw, or of its first tick. This gives a uniform random permutation. At the first activation of a process $v$, its same-colored neighbors of smaller label have been activated earlier, while $v$ still held its initial color, hence in conflict, and have moved away from that color, while those of larger label still hold it. So $v$ moves at its first activation iff it is not a peak, and never again by Lemma 1(1). The configuration becomes legitimate when the last non-peak is activated for the first time, since every initially conflicting edge has a non-peak endpoint and moves never create conflicts. In model (1) the gaps between successive first draws are independent of the order in which the processes first appear, so given that order the number of steps until the $J$-th distinct process is drawn has mean $\sum_{i=0}^{J-1} n/(n-i) = n(H_n - H_{n-J})$. The last process drawn carries the largest label and is always a peak, so $J \le n-1$, with $J = n-1$ on the monochrome clique, where every other process is a non-peak. ◀

▶ **Proposition 2** (restated). *For an initial configuration* $\gamma$ *with conflict graph* $G_\gamma$*: (1)* $T(G,\gamma) = n - S(G_\gamma)$ *in law, hence* $\mathbb{E}\, T(G,\gamma) = \sum_v d_v(\gamma)/(d_v(\gamma)+1)$*; (2)* $S(G_\gamma) \ge S(G)$ *for every labeling; (3) over all central and locally central daemons the number of moves takes exactly the values* $n - \alpha(G_\gamma), \ldots, n - \kappa(G_\gamma)$.

**Proof of Proposition 2.** (1) Under $\gamma$ two neighbors are in conflict exactly when they are adjacent in $G_\gamma$, and a peak is a local maximum of $G_\gamma$, so this is Lemma 1 with the definition unfolded, and the mean follows from $\mathbb{P}(v \text{ is a peak}) = 1/(d_v(\gamma)+1)$. (2) A vertex whose label exceeds those of its neighbors in $G$ exceeds those of its neighbors in the subgraph $G_\gamma$. (3) Every execution under a central daemon is the smallest-label schedule of some labeling: label the processes activated in the execution $1, \ldots, T$ in activation order and the others arbitrarily above $T$. By Lemma 1(1) an activated process is never enabled again, so at step $t$ the enabled process of smallest label is the one labeled $t$, and the execution has $n - S(G_\gamma)$ moves for that labeling. A step of a locally central daemon activates pairwise non-adjacent processes and can be serialized in any order without changing the moves, so the same holds for these daemons. Conversely every labeling yields an execution. Local maxima form an independent set meeting every component (each component contains its largest label), so $\kappa(G_\gamma) \le S(G_\gamma) \le \alpha(G_\gamma)$, and every independent set $I$ meeting every component is the set of local maxima of some labeling: give the largest labels to $I$ and let the other labels decrease with the distance to $I$. Removing vertices one at a time from a maximum independent set, inside the components where it has at least two, gives such sets of every size between $\alpha(G_\gamma)$ and $\kappa(G_\gamma)$. Finally $\kappa(G_\gamma) \geq \kappa(G)$, with equality for the monochrome configuration. ◀

▶ **Theorem 1** (restated). *Under a uniform $k$-coloring the expected number of peaks is given by Eq. (1).*

**Proof of Theorem 1.** By linearity, $\mathbb{E}(S_n) = n \cdot \mathbb{P}(v \text{ is a peak})$ for a uniformly random vertex $v$. Condition on the degree of $v$ being $d$. The $d$ neighbors of $v$ each share $v$'s color independently with probability $\frac{1}{k}$, so the number $m$ of same-colored neighbors is $\mathsf{Binomial}(d, \frac{1}{k})$. Given $m$, the labels of $v$ and these $m$ neighbors are in uniform random relative order, so $v$ is the maximum with probability $\frac{1}{m+1} = \int_0^1 t^m \, \mathrm{d}t$. Hence

$$\begin{aligned} \mathbb{P}(v \text{ is a peak} \mid d) &= \int_0^1 \mathbb{E}[t^m] \, \mathrm{d}t = \int_0^1 \left(\frac{k-1+t}{k}\right)^d \mathrm{d}t \\ &= k \int_{1-1/k}^1 u^d \, \mathrm{d}u = \frac{k}{d+1}\left(1 - \left(1 - \frac{1}{k}\right)^{d+1}\right). \end{aligned} \tag{25}$$

Averaging over $d$ with weights $p_d$ and multiplying by $n$ gives Eq. (1). ◀

▶ **Theorem 2** (restated). *In the monochrome case, the two vertices of an ordered non-adjacent pair with parameters $(a, b, c)$ are both peaks with probability*

$$J(a,b,c) = \frac{1}{a+b+c+2}\left(\frac{1}{a+b+1} + \frac{1}{b+c+1}\right),$$

*and those of an adjacent pair never are. With $p_{a,b,c}$ the probability that a random ordered pair is non-adjacent with parameters $(a, b, c)$,*

$$\mathbb{E}(S_n(S_n - 1)) = n(n-1) \sum_{a,b,c} \frac{p_{a,b,c}}{a+b+c+2}\left(\frac{1}{a+b+1} + \frac{1}{b+c+1}\right).$$

**Proof of Theorem 2.** Assign i.i.d. uniform $[0, 1]$ ranks. Since the peak condition depends only on relative order, these have the same law as a uniform random labeling. Condition on the ranks $s, t$ of $u, v$: each of the $a$ private neighbors of $u$ must rank below $s$, each of the $c$ private neighbors of $v$ below $t$, and each of the $b$ common neighbors below $\min(s, t)$, so $\mathbb{P}(u, v \text{ both peaks}) = \int_0^1 \int_0^1 s^a t^c \min(s,t)^b \, \mathrm{d}s \, \mathrm{d}t$, and splitting on $s \leq t$ and $s \geq t$ evaluates this to $J(a, b, c)$. Summing over pairs with weights $p_{a,b,c}$ and using Lemma 1 gives Eq. (3). ◀

▶ **Theorem 3** (restated). *The joint peak probabilities of a pair with parameters $(a, b, c)$ under a uniform $k$-coloring are the double integrals $\pi_{\text{same}}$ and $\pi_{\text{dist}}$ of Eq. (4), and the second factorial moment is given by Eq. (5).*

**Proof of Theorem 3.** Realize the labeling by i.i.d. uniform ranks and condition on the ranks $x, y$ and on the colors of $u$ and $v$. A neighbor of $u$ prevents $u$ from being a peak iff it has the color of $u$ and a rank above $x$, an event of probability $(1 - x)/k$, independently across neighbors, so each private neighbor of $u$ leaves $u$ a peak with probability $s = (k - 1 + x)/k$, and each private neighbor of $v$ leaves $v$ a peak with probability $t = (k - 1 + y)/k$. A common neighbor leaves both $u$ and $v$ peaks with probability $\min(s, t)$ when $u$ and $v$ share a color (it needs another color, or a rank below both) and $s + t - 1$ when they do not (it must avoid two disjoint events of probabilities $(1 - x)/k$ and $(1 - y)/k$). Given the colors and the ranks, both are therefore peaks with probability $s^a t^c \min(s, t)^b$ in the first case and $s^a t^c (s + t - 1)^b$ in the second. As $x, y$ are uniform on $[0, 1]$, the variables $s, t$ are uniform on $[1 - 1/k, 1]$ with density $k$ each, and the two cases have probabilities $1/k$ and $(k - 1)/k$: this is Eq. (4). Two adjacent vertices of the same color are never both peaks, and adjacency does not enter the second case, so adjacent pairs contribute $\pi_{\text{dist}}$ only. Summing over ordered pairs with the weights $p_{a,b,c}$ and $q_{a,b,c}$ gives Eq. (5). In the original variables, with $s = (k-1+xy)/k \le t = (k-1+y)/k$ parametrizing the half of the square where $s \le t$ (Jacobian $y/k^2$) and $s = (k-1+x)/k$, $t = (k-1+y)/k$ in $\pi_{\text{dist}}$,

$$\pi_{\text{same}}(a,b,c) = \frac{1}{k}\int_0^1\int_0^1 y\left[\left(\frac{k-1+xy}{k}\right)^{a+b}\left(\frac{k-1+y}{k}\right)^{c} + \left(\frac{k-1+xy}{k}\right)^{b+c}\left(\frac{k-1+y}{k}\right)^{a}\right]\mathrm{d}x\,\mathrm{d}y, \tag{26}$$

$$\pi_{\text{dist}}(a,b,c) = \frac{k-1}{k}\int_0^1\int_0^1 \left(\frac{k-1+x}{k}\right)^{a}\left(\frac{k-1+y}{k}\right)^{c}\left(\frac{x+y+k-2}{k}\right)^{b}\mathrm{d}x\,\mathrm{d}y,$$

the forms used in the random-graph computations of Appendix E and Appendix F. ◀

▶ **Corollary 7** (Explicit finite sums). *For $k \ge 2$ the integrals of Eq. (26) evaluate to*

$$\pi_{\text{same}}(a,b,c) = \frac{1}{k^{a+b+c+1}}\left[\sum_{i=0}^{a+b}\sum_{j=0}^{c}\binom{a+b}{i}\binom{c}{j}\frac{(k-1)^{a+b+c-i-j}}{(i+1)(i+j+2)} + (a \leftrightarrow c)\right],$$

$$\pi_{\text{dist}}(a,b,c) = \frac{\left(\frac{k-1}{k}\right)^{a+c+1}}{k^b}\sum_{j=0}^{b}\sum_{i=0}^{j}\binom{b}{j}\binom{j}{i}(k-2)^{b-j}K_a(i)K_c(j-i), \text{with} \tag{27}$$

$$K_a(i) = \sum_{m=0}^{a}\binom{a}{m}\frac{1}{(k-1)^m(i+m+1)},$$

*where $(a \leftrightarrow c)$ is the first double sum with $a$ and $c$ exchanged.*

**Proof.** Expand the binomials in Eq. (26) and integrate term by term. (At $k = 2$ the factor $(k-2)^{b-j}$ is read as $0^0 = 1$ when $j = b$.) ◀

▶ **Corollary 2** (restated). *On a graph of maximum degree $\Delta$, $\mathbb{P}(|T - \mathbb{E}\,T| \ge t) \le 2\exp(-2t^2/(n(\Delta+1)^2))$, and the Kolmogorov distance between the standardized stabilization time and the normal law is at most $C\Delta^4 n/\mathrm{Var}(T)^{3/2}$.*

**Proof of Corollary 2.** Realize the labeling by i.i.d. uniform ranks $U_v$ and the colors by independent variables $C_v$, and group $(U_v, C_v)$ into one variable per vertex. The peak count is a function of these $n$ independent variables, and changing one of them changes the peak status of $v$ and of its neighbors only, hence by at most $\Delta + 1$, so (1) is the bounded differences inequality [McD89]. For (2), the indicator that $v$ is a peak depends only on the variables of $v$ and its neighbors, so two families of indicators whose vertices are pairwise at distance at least 3 depend on disjoint sets of independent variables and are independent. The graph joining pairs at distance at most 2, of maximum degree at most $\Delta^2$, is thus a dependency graph for the indicators, and the normal approximation theorem for sums of bounded variables with a dependency graph [BR89, Rin94] bounds the Kolmogorov distance by a constant times $nD^2/\sigma^3 + \sqrt{n}D^{3/2}/\sigma^2 + D/\sigma$ with $D = \Delta^2 + 1$ (one plus the maximum degree of the dependency graph) and $\sigma^2 = \mathrm{Var}(T)$. Since $\sigma^2 \le n(\Delta+1)^2/2$ by the Efron-Stein inequality applied to the same $n$ variables, and $(\Delta+1)^2 \le 2D$, the first term dominates the other two, and $D^2 \le 4\Delta^4$. The bound tends to 0 exactly when $\sigma^2/n^{2/3} \to \infty$. For a slower growth of the variance it says nothing, and we do not claim normality there. ◀

On the path the second factorial moment reproduces $\mathrm{Var}\big(T_{P_n}\big) = (2n+2)/45$ for $n \ge 4$.

## B Proofs for Section 4

**Randomized analogues of the extremal graphs.** On a random graph the mean simply averages over the degree law (Theorem 1), and the extremal graphs of Section 4 have immediate randomized analogues: a uniform random perfect matching for the minimal mean, a uniform random partition of the vertices into a *fixed* number $\kappa$ of cliques for zero variance (the stabilization time stays $n-\kappa$, so it is still deterministic even though the graph is genuinely random once $\kappa \geq 2$), and a star with a uniformly random center for the maximal variance.

▶ **Proposition 3** (restated). *Among graphs on $n$ vertices without isolated vertices, $\mathbb{E}(T)$ ranges from $n/2$ (perfect matching, $n$ even) or $(3n+1)/6$ ($n$ odd) to $n-1$ (clique). The extremal graphs are the same for every $k$.*

**Proof of Proposition 3.** From $\mathbb{E}(T) = n - \sum_v 1/(d_v+1)$ and the strict decrease of $1/(d+1)$ in $d$: every degree obeys $1 \leq d_v \leq n-1$, so $1/n \leq 1/(d_v+1) \leq 1/2$, and summing gives $1 \leq \sum_v 1/(d_v+1) \leq n/2$. The right bound is met iff every $d_v = 1$ (the perfect matching, $n$ even), whence $\mathbb{E}(T) = n/2$, and the left bound iff every $d_v = n-1$ (the clique $K_n$), whence $\mathbb{E}(T) = n-1$. For odd $n$ at least one vertex has degree at least 2, since the number of odd-degree vertices is even, so $\sum_v 1/(d_v+1) \leq (n-1)/2 + 1/3$, with equality for a path on three vertices plus a perfect matching of the other vertices. For general $k$ the same termwise argument applies to $\mathbb{E}(T) = n - \sum_v \rho_k(d_v)$, because $\rho_k$ is strictly decreasing: $\rho_k(d) - \rho_k(d+1) = k\int_{1-1/k}^{1} u^d(1-u)\,\mathrm{d}u > 0$. ◀

▶ **Theorem 4** (restated). *Monochrome,* $\mathrm{Var}(T) = 0$ *iff $G$ is a disjoint union of cliques.*

**Proof of Theorem 4.** If $G$ is a disjoint union of cliques then, since every labeling of a clique has a unique local maximum, each clique contributes exactly one peak under *any* labeling, so $S = \kappa(G)$ and $T = n - \kappa(G)$ are constant (isolated vertices are cliques of size 1, each its own component). Conversely, a graph is a disjoint union of cliques iff it contains no induced path on three vertices. So if $G$ is not such a union it has vertices $a, b, c$ with $ab, bc \in E$ and $ac \notin E$. Fix any assignment of the $n-3$ smallest labels to $V \setminus \{a, b, c\}$, and give $a, b, c$ the three largest labels. The peak status of every vertex outside $\{a, b, c\}$ is then identical in all such labelings: such a vertex ranks below $a, b, c$, hence is a non-peak if adjacent to any of them and otherwise sees only the fixed outside labels. Among $\{a, b, c\}$, $a$ is a peak iff $a > b$ (its only possibly-larger neighbor), likewise $c$ iff $c > b$, and $b$ iff it exceeds both $a$ and $c$. Ranking $b$ below both $a$ and $c$ gives 2 peaks, and ranking $b$ above both gives 1. The two labelings are equally likely and give peak counts differing by 1, so $S$ is non-constant and $\mathrm{Var}(T) > 0$. ◀

▶ **Proposition 4** (restated). *On the monochrome star, $\mathbb{P}(T = j) = 1/n$ for $j \in \{1, \ldots, n-2\}$ and $\mathbb{P}(T = n-1) = 2/n$, with the mean and variance of Eq. (7).*

**Proof of Proposition 4.** By Lemma 1, $T = n - S$. On the star a leaf is a peak iff its label exceeds the center's (its only neighbor), and the center is a peak iff it holds the maximum label. Let $c$ be the label rank of the center. For $c < n$ the $n-c$ leaves above the center are the peaks, so $T = c$. For $c = n$ the center is the unique peak and $T = n-1$. As $c$ is uniform on $\{1, \ldots, n\}$, the value $n-1$ arises from both $c = n-1$ and $c = n$, giving the stated law. The moments and their expansion are a direct computation. ◀

▶ **Lemma 2** (restated). *For distinct vertices $u, v$ with peak indicators $X_u, X_v$,* $\mathrm{Cov}(X_u, X_v) \leq 1/12$*, with the finer classification stated in the body.*

**Proof of Lemma 2.** Adjacent vertices are never both peaks, so there $\mathrm{Cov}(X_u, X_v) = -\,\mathbb{E}[X_u]\,\mathbb{E}[X_v] < 0$. For non-adjacent $u, v$ with parameters $(a, b, c)$ $(a = d_u - b,\ c = d_v - b)$, draw i.i.d. uniform $[0,1]$ values (equivalent to a random labeling) and condition on the values $s, t$ of $u, v$: a private neighbor of $u$ must fall below $s$, one of $v$ below $t$, and a common neighbor below $\min(s,t)$, so

$$\mathbb{E}[X_u X_v] = \int_0^1 \int_0^1 s^a t^c \min(s,t)^b \,\mathrm{d}s\,\mathrm{d}t = \frac{1}{a+b+c+2}\left(\frac{1}{a+b+1} + \frac{1}{b+c+1}\right), \quad (28)$$

the square being split along its diagonal: on $t < s$ the inner integral is $s^{a+b+c+1}/(b+c+1)$, on $s < t$ it is $t^{a+b+c+1}/(a+b+1)$. Subtracting $\mathbb{E}[X_u]\,\mathbb{E}[X_v] = 1/((a+b+1)(b+c+1))$ over the common denominator gives Eq. (8), the numerator being $(a+b+1) + (b+c+1) - (a+b+c+2) = b$. The rest is read off Eq. (8). The covariance vanishes with $b$. For $b \geq 1$ the three factors of the denominator are at least $b+1$, $b+1$ and $b+2$, with equality iff $a = c = 0$, whence the bound $b/((b+1)^2(b+2))$, which is $1/12$ at $b = 1$, $1/18$ at $b = 2$ and decreases afterwards (there $b(b+2)(b+3) > (b+1)^3$). For $b = 1$ and $(a, c) \neq (0, 0)$ one of the first two factors is at least 3 and the third at least 4, whence $1/24$. For $(a, b) = (0, 1)$ the denominator is $2(c+2)(c+3)$. Equality in $\mathrm{Cov}(X_u, X_v) \leq 1/12$ thus requires $b = 1$ and $a = c = 0$: two leaves with the same single neighbor. ◀

▶ **Theorem 6** (restated). *For every $n \geq 3$, among all graphs on $n$ vertices,* $\mathrm{Var}(T)$ *is maximized by the star* $\mathrm{St}_n$ *and by no other graph.*

**Proof of Theorem 6.** We first prove the statement among graphs without isolated vertices, for every $n \geq 2$. For $n = 2$ the only such graph is the edge $\mathrm{St}_2$, and for $3 \leq n \leq 6$ the statement was checked by exhaustive enumeration of all labeled graphs without isolated vertices (Appendix G). Let $n \geq 7$ and let $G$ be a graph on $n$ vertices without isolated vertices, other than $\mathrm{St}_n$. Call a vertex of degree 1 a *leaf* and its unique neighbor its *hub*. Write $\ell_w$ for the number of leaves at $w$, $L = \sum_w \ell_w$ for the number of leaves, $x = n - L$ for the number of vertices of degree at least 2, and $\ell^* = \max_w \ell_w \leq L$. The star is the graph with $\ell^* = n - 1$. If $x = 0$ then $G$ is a perfect matching and $\mathrm{Var}(T) = 0$ by Theorem 4, so let $x \geq 1$. We bound $\mathrm{Var}(T) = \sum_v \mathrm{Var}(X_v) + \sum_{u \neq v} \mathrm{Cov}(X_u, X_v)$ class by class with Lemma 2, writing $p_v = 1/(d_v + 1)$ for $\mathbb{E}(X_v)$.

*Diagonal.* $\mathrm{Var}(X_v) = p_v(1 - p_v)$ is $1/4$ for a leaf and at most $2/9$ for a vertex of degree at least 2: at most $L/4 + 2x/9$ in all.

*Two leaves.* Two leaves at the same hub have parameters $(0, 1, 0)$ and covariance $1/12$. Two leaves at distinct hubs have no common neighbor and covariance 0, unless they form an isolated edge, with negative covariance. The ordered pairs of leaves at a common hub number $\sum_w \ell_w(\ell_w - 1) \leq L(\ell^* - 1)$.

*A leaf and a vertex of degree at least 2.* Let $u$ be a leaf at $w$ and $v \neq w$ a vertex of degree at least 2. The pair is non-adjacent, and has a common neighbor iff $v$ is adjacent to $w$. Its parameters are then $(0, 1, d_v - 1)$ and its covariance $1/(2(d_v+1)(d_v+2)) \leq 1/24$. Every neighbor of $w$ that is not a leaf at $w$ has degree at least 2, so these unordered pairs number $\sum_w \ell_w(d_w - \ell_w) \leq L(x - 1)$ (a hub of degree at least 2 is one of the $x$ vertices and has at most $x - 1$ such neighbors, and a hub of degree 1 has none), and they contribute at most $L(x-1)/12$ to the sum over ordered pairs.

*Two vertices of degree at least 2.* If non-adjacent with a common neighbor, their covariance is at most $1/18$ by Lemma 2 (they are not two leaves at the same hub). The ordered non-adjacent pairs among these $x$ vertices number $x(x-1) - 2e'$, where $e'$ counts the edges between them. A vertex of degree at least 2 carrying no leaf has at least two such edges and at most $L$ vertices carry a leaf, so $2e' \geq 2(x - L)^+$.

*Adjacent pairs* have negative covariance, and we keep only the two ordered pairs formed by a leaf and its hub $w$, which contribute $-2p_u p_w = -p_w \leq -1/n$ together when $w$ has degree at least 2. The two vertices of an isolated edge are each other's hub, and the edge contributes $-1/2 \leq -2/n$ for its two leaves once $n \geq 4$. In all, the kept pairs contribute at most $-L/n$. Adding up,

$$\mathrm{Var}(T) \leq \frac{L}{4} + \frac{2x}{9} + \frac{L(\ell^* - 1)}{12} + \frac{L(x-1)}{12} + \frac{x(x-1) - 2(x-L)^+}{18} - \frac{L}{n}, \tag{29}$$

to be compared with $\mathrm{Var}_{\mathrm{St}_n}(T) = (n-1)/4 + (n-1)(n-2)/12 - (n-1)/n + (n-1)/n^2$, the value of the same decomposition on the star (it is Eq. (7)).

*Case* $x = 1$. Then $L = n - 1$ and $\ell^* \leq n - 2$ since $G$ is not the star, and Eq. (29) gives $\mathrm{Var}(T) \leq \mathrm{Var}_{\mathrm{St}_n}(T) + 2/9 - (n-1)/n^2 - (n-1)/12 < \mathrm{Var}_{\mathrm{St}_n}(T)$.

*Case* $x \geq 2$. The right-hand side of Eq. (29) increases with $\ell^*$, so $\ell^*$ may be replaced by $L = n - x$, after which $\mathrm{Var}_{\mathrm{St}_n}(T)$ minus the right-hand side equals

$$D(x) = \frac{x-1}{4} - \frac{2x}{9} + \frac{(x-1)(n-2)}{12} - \frac{x(x-1) - 2(2x-n)^+}{18} - \frac{x-1}{n} + \frac{n-1}{n^2}. \tag{30}$$

As a function of a real variable, $D$ is concave on $[2, n/2]$ and on $[n/2, n]$ (its only non-linear terms are $-x(x-1)/18$ and $(2x-n)^+/9$, and the second is convex, which is why the interval is split), so on each interval it is at least the smaller of its values at the endpoints:

$$\begin{aligned} D(2) &= \frac{3n-17}{36} - \frac{1}{n^2}, \\ D(n/2) &= \frac{(n-2)(n-3)}{36} + \frac{n-18}{72} - \frac{n-2}{2n} + \frac{n-1}{n^2}, \\ D(n) &= \frac{(n-3)(n+1)}{36} - \frac{(n-1)^2}{n^2}. \end{aligned} \tag{31}$$

At $n = 7$ these are $4/36 - 1/49$, $0.168...$ and $32/36 - 36/49$, all positive. For $n \geq 8$, $D(2)$ increases with $n$ from $7/36 - 1/64$, $D(n/2) \geq (n-2)(n-3)/36 + (n-18)/72 - 1/2 \geq 30/36 - 10/72 - 1/2$ and $D(n) \geq (n-3)(n+1)/36 - 1 \geq 1/4$. Hence $D(x) > 0$ for every integer $x \in \{2, ..., n\}$, that is $\mathrm{Var}(T) < \mathrm{Var}_{\mathrm{St}_n}(T)$.

*Isolated vertices.* Let $n \geq 3$ and let $G$ have $i \geq 1$ isolated vertices. They are peaks of every labeling, so $\mathrm{Var}(T)$ equals the variance on the graph $G'$ obtained by deleting them, which has no isolated vertex: either $G'$ is empty and $\mathrm{Var}(T) = 0$, or $G'$ has $n - i \geq 2$ vertices and $\mathrm{Var}(T) \leq \mathrm{Var}_{\mathrm{St}_{n-i}}(T)$ by the above. By Eq. (7), $\mathrm{Var}_{\mathrm{St}_n}(T) = (n^2-1)/12 - (n-2)/n - 1/n^2$, which increases strictly with $n \geq 2$ (its increments are $(2n+1)/12 - 2/(n(n+1)) + 1/n^2 - 1/(n+1)^2 > 0$), so $\mathrm{Var}(T) < \mathrm{Var}_{\mathrm{St}_n}(T)$ in both cases. ◀

The companion code checks this result on various families of graphs, including all graphs on $n \leq 7$ vertices.

## C Path and cycle: proofs and derivations

**Degree-driven means.** The path and cycle are two cases of a general pattern: by Theorem 1 the mean depends on the graph through its degrees alone, so it is a simple closed form the moment the degree sequence is known. For a graph on $n$ vertices with $\ell$ leaves and $n - \ell$ vertices of degree $d$, the monochrome mean is

$$\mathbb{E}(T) = \frac{\ell}{2} + (n-\ell)\frac{d}{d+1}, \tag{32}$$

and for general $k$ the two terms become $\ell(1 - \rho_k(1))$ and $(n-\ell)(1 - \rho_k(d))$. Taking $\ell = 0$ gives the $d$-regular graphs. The cycle is $(\ell, d) = (0, 2)$ and the path is $(\ell, d) = (2, 2)$, while the $d$-regular trees have $\ell > 0$ (though $\ell > 0$ does not force a tree). The variance is not degree-driven this way: it depends on the joint degrees of neighboring vertices through Theorem 3, so no comparable closed form holds for it.

**Proof of Corollary 4.** By Eq. (6) only the pairs at distance one and two contribute. On the cycle with $n \geq 5$ every vertex is a peak with probability $\rho = \rho_k(2)$, its two neighbors form with it ordered pairs with parameters $(1, 0, 1)$ and its two vertices at distance two pairs with parameters $(1, 1, 1)$, so

$$\mathrm{Var}\left(T_{C_n}\right) = n[\rho(1-\rho) + 2(\pi_{\text{dist}}(1,0,1) - \rho^2) + 2((\pi_{\text{same}} + \pi_{\text{dist}})(1,1,1) - \rho^2)], \quad (33)$$

and evaluating the integrals of Eq. (4) at these parameters gives the stated coefficient. On the path with $n \geq 4$ the two endpoints are peaks with probability $\rho_k(1)$, the four ordered adjacent pairs they belong to have parameters $(0, 0, 1)$ or $(1, 0, 0)$ and the four ordered pairs at distance two $(0, 1, 1)$ or $(1, 1, 0)$. The $2(n-3)$ other adjacent pairs and $2(n-4)$ other pairs at distance two are as on the cycle. The difference with the cycle is the stated boundary term. At $k = 1$, $\rho = 1/3$, $\pi_{\text{dist}} = 0$ and $\pi_{\text{same}}(1,1,1) = 2/15$, so $\mathrm{Var}\left(T_{C_n}\right)/n = 2/9 - 2/9 + 2(2/15 - 1/9) = 2/45$. The monochrome moments of the path are classical [Hiv+07, WS96]. ◀

**Proof of Theorem 7.** *Monochrome path.* Removing the smallest value of a permutation of size at least 2 splits it either at an end (leaving one permutation) or in the interior (leaving two independent ones). The removed element is never a peak, and its neighbors beat it, so their peak status is unchanged once they become endpoints of the pieces, while the permutation of size 1 is a peak. This gives the Riccati equation $\partial_z P = v + 2P + P^2$ with $P(0, v) = 0$, whose solution is Eq. (11) [Ent69, Zhu16]. The probability generating function of $T_n$ follows by marking non-peaks, $\mathbb{E}(u^{T_n}) = u^n[z^n]P(z, 1/u)$. Coefficient extraction gives $\mathbb{E}(T_n) = (2n-1)/3$ and, for $n \geq 4$, $\mathrm{Var}(T_n) = (2n+2)/45$, in agreement with Corollary 3 and Corollary 4.

*$k$-colored path.* Under a uniform $k$-coloring a vertex is a peak when it exceeds its *same-color* neighbors only, so the path breaks at every color change into maximal monochromatic runs whose peaks are independent. Each run is a monochrome path, counted by $P(z, v)$, and consecutive runs carry distinct colors. Summing over the number of runs gives $kP/(1 - (k-1)P)$, whose value at $v = 1$ is $kz/(1 - kz)$, i.e. $[z^n/n!] = n!k^n$, the number of labeled $k$-colored paths.

*Monochrome oriented cycle.* The cycle of length 1 is a loop and its vertex is not a peak, by convention. In an oriented cycle of size at least 2, removing the minimum-labeled vertex, which is never a peak, produces an oriented path whose endpoints keep their peak status, hence $Y(z, v) = z + \int_0^z P(y, v)\,\mathrm{d}y$, which integrates to the closed form of Eq. (12). In particular the peak count of $C_{n+1}$ has the law of that of $P_n$, which transfers the monochrome moments and the central limit theorem of the path to the cycle. The mean $n/3$ and variance $2n/45$ (for $n \geq 5$) of the series are those of $S_n = n - T_n$ in Corollary 3 and Corollary 4.

*$k$-colored oriented cycle.* Let $P_{\text{same}}(z, v)$ (resp. $P_{\text{dist}}(z, v)$) count $k$-colored paths whose first and last vertices have the same (resp. distinct) colors. Any path is in one of the two cases, and a path with same-color ends is either monochrome, or decomposes uniquely as a path with distinct-color ends followed by a monochrome path of the color of its starting point, so

$$P_{\text{same}} + P_{\text{dist}} = \frac{kP}{1 - (k-1)P}, \qquad P_{\text{same}} = kP + P_{\text{dist}}P, \quad (34)$$

hence

$$P_{\text{same}} = \frac{(1-(k-2)P)kP}{(1+P)(1-(k-1)P)}, \qquad P_{\text{dist}} = \frac{k(k-1)P^2}{(1+P)(1-(k-1)P)}. \tag{35}$$

A $k$-colored oriented cycle of size 1 contributes $kz$. Otherwise, removing the vertex of label 1 leaves a $k$-colored path whose ends keep their peak status, and the removed vertex is a peak in the cycle iff its color differs from the colors of both ends of this path: when the ends share a color, $k-1$ colors make it a peak and one does not, whereas when they differ, $k-2$ do and two do not. This is Eq. (13), which collapses to $Y$ at $k=1$. The cycles of length 1 and 2, a loop and a double edge, are outside the simple-graph model. Their terms are the normalizations that make the differential equations hold, and the series describe $C_n$ for $n \geq 3$. ◀

The companion code checks all four series.

## D Random trees: proofs and derivations

The companion code checks the moments of Corollary 6.

**The $k$-colored mean.** By Theorem 1, $\mathbb{E}(S_n) = n\,\mathbb{E}[\rho_k(D)]$ with $D$ the degree of a uniform vertex and $\rho_k(d) = k\int_{1-1/k}^{1} t^d\,\mathrm{d}t$. The number of labeled trees on $n$ vertices in which a given vertex has degree $d$ is $\binom{n-2}{d-1}(n-1)^{n-1-d}$, so $D-1 \sim \text{Binomial}(n-2, 1/n)$ and $\mathbb{E}[t^D] = t((n-1+t)/n)^{n-2}$, which is Eq. (17). Since $((n-1+t)/n)^{n-2} = e^{t-1}(1+O(1/n))$ uniformly on $[0,1]$ and $k\int_{1-1/k}^{1} te^{t-1}\,\mathrm{d}t = e^{-1/k}$, the mean is $\left(1-e^{-1/k}\right)n + O(1)$. In closed form, with $m = n-2$, $a = (n-1)/n$ and $b = 1/n$, the integral of Eq. (17) is

$$\int_{1-1/k}^{1} t(a+bt)^m\,\mathrm{d}t = \left[\frac{t(a+bt)^{m+1}}{b(m+1)} - \frac{(a+bt)^{m+2}}{b^2(m+1)(m+2)}\right]_{1-1/k}^{1}, \tag{36}$$

so that $\mathbb{E}(S_n) = nk$ times this bracket. At $k=1$ the bracket equals $(1-1/n)^{n-1}$ and the formula reduces to $\mathbb{E}(S_n) = n(1-1/n)^{n-1}$.

**The $k$-colored variance.** By Corollary 1 only the pairs at distance at most two contribute to the covariance, and in a tree the neighborhood profile of such a pair is read off the degrees of its two vertices: with the parameters $(a,b,c)$ of Section 3, $b \leq 1$, and $(a,c) = (d_u - 1, d_v - 1)$ when $uv$ is an edge or when $u$ and $v$ are at distance two (then $b = 1$), $(a,c) = (d_u, d_v)$ beyond.

▶ **Lemma 3** (Pairs of vertices in a Cayley tree). *In the uniform random labeled tree on $n \geq 4$ vertices, for two fixed distinct vertices $u, v$ and all real $s, t$:*

1. $\mathbb{E}[s^{d_u} t^{d_v}] = st((s+t+n-2)/n)^{n-2}$*: the vector $(d_u - 1, d_v - 1)$ is multinomial with $n-2$ trials and probabilities $1/n, 1/n$.*
2. *$u$ and $v$ are adjacent with probability $2/n$, and then* $\mathbb{E}[s^a t^c] = \frac{s+t}{2}((s+t+n-2)/n)^{n-3}$*: $(a,c)$ is the sum of a multinomial vector with $n-3$ trials and probabilities $1/n, 1/n$ and of a vector uniform on $\{(1,0),(0,1)\}$.*
3. *$u$ and $v$ are at distance two with probability $3(n-2)/n^2$, and then* $\mathbb{E}[s^a t^c] = \frac{s+t+1}{3}((s+t+n-2)/n)^{n-4}$*: $(a,c)$ is the sum of a multinomial vector with $n-4$ trials and probabilities $1/n, 1/n$ and of a vector uniform on $\{(1,0),(0,1),(0,0)\}$.*

**Proof.** Let $C(z)$ be the Cayley tree function of Section 5.2, with $C(z) = ze^{C(z)}$, the exponential generating function of rooted labeled trees. A tree with an ordered pair of distinct distinguished vertices $u, v$ is the path $u = x_0, x_1, \ldots, x_k = v$ $(k \geq 1)$ together with a rooted tree hanging from each $x_i$. Mark by $s$ (resp. $t$) the children of the root of the tree hanging from $u$ (resp. $v$), that is, the $d_u - 1$ (resp. $d_v - 1$) neighbors of $u$ (resp. $v$) off the path. A rooted tree whose root has its children marked by $s$ has generating function $ze^{sC(z)}$. Summing over all $k \geq 1$, keeping $k = 1$ only, or keeping $k = 2$ only gives, for all pairs, adjacent pairs and pairs at distance two,

$$A_1(z) = \frac{z^2 e^{(s+t)C(z)}}{1 - C(z)}, \quad A_2(z) = z^2 e^{(s+t)C(z)}, \quad A_3(z) = z^2 C(z) e^{(s+t)C(z)}, \tag{37}$$

and in the last two cases $(d_u - 1, d_v - 1) = (a, c)$. Since $z = C(z)e^{-C(z)}$, each $A_i(z)$ is $H_i(C(z))/(1 - C(z))$ with $H_1(u) = u^2 e^{(s+t-2)u}$, $H_2(u) = (1-u)u^2 e^{(s+t-2)u}$ and $H_3(u) = (1-u)u^3 e^{(s+t-2)u}$. The Lagrange inversion formula $[z^n]H(C(z))/(1 - C(z)) = [u^n]H(u)e^{nu}$ [FS09] then gives

$$\begin{aligned} n!\,[z^n]A_1(z) &= n(n-1)(s+t+n-2)^{n-2}, \\ n!\,[z^n]A_2(z) &= n(n-1)(s+t)(s+t+n-2)^{n-3}, \\ n!\,[z^n]A_3(z) &= n(n-1)(n-2)(s+t+1)(s+t+n-2)^{n-4}. \end{aligned} \tag{38}$$

All pairs play the same role, and there are $n(n-1)n^{n-2}$ trees with an ordered pair of distinct vertices. Dividing by this number gives $\mathbb{E}[s^{d_u-1}t^{d_v-1}]$, hence (1), and $\mathbb{E}[s^a t^c]$ restricted to adjacent pairs and to pairs at distance two. At $s = t = 1$ these give the probabilities $2/n$ and $3(n-2)/n^2$, and dividing by them gives the conditional expectations of (2) and (3). The descriptions as a multinomial vector plus a uniform vector are read off these products.◀

▶ **Proposition 10** (Random tree, variance under a $k$-coloring). *Write* $\langle f \rangle = k^2 \int_{1-1/k}^1 \int_{1-1/k}^1 f(s,t)\,\mathrm{d}s\,\mathrm{d}t$ *and* $\varphi = (s+t+n-2)/n$. *Under a uniform random* $k$*-coloring of the uniform random labeled tree on* $n \geq 4$ *vertices,*

$$\begin{aligned} \mathbb{E}(S_n(S_n - 1)) = n(n-1)\langle st\varphi^{n-2}\rangle &+ 2(n-1)\langle \frac{s+t}{2}\varphi^{n-3}\left(\frac{k-1}{k} - st\right)\rangle \\ &+ \frac{3(n-1)(n-2)}{n}\langle \frac{s+t+1}{3}\varphi^{n-4}\left(\frac{\min(s,t)}{k} + \frac{k-1}{k}(s+t-1) - st\right)\rangle, \end{aligned} \tag{39}$$

*and* $\mathrm{Var}(T_n) = \mathbb{E}(S_n(S_n-1)) + \mathbb{E}(S_n) - \mathbb{E}(S_n)^2$ *with* $\mathbb{E}(S_n)$ *given by Eq. (17). As* $n \to \infty$, $\mathrm{Var}(T_n) = \sigma_k^2 n + O(1)$ *with* $\sigma_k^2 = e^{-1/k} - (1 + k^{-2})e^{-2/k}$*, as stated in Eq. (18).*

**Proof.** By Theorem 3, in the variables of Eq. (4), two vertices with parameters $(a, b, c)$ are both peaks with probability $\langle s^a t^c \min(s,t)^b\rangle/k + (k-1)\langle s^a t^c (s+t-1)^b\rangle/k$ if non-adjacent and $(k-1)\langle s^a t^c (s+t-1)^b\rangle/k$ if adjacent, while $\rho_k(a)\rho_k(c) = \langle s^a t^c\rangle$. Split the ordered pairs of distinct vertices into adjacent pairs ($b = 0$), pairs at distance two ($b = 1$) and the others, which are both peaks with probability $\rho_k(d_u)\rho_k(d_v)$ by Corollary 1. Summing $\rho_k(d_u)\rho_k(d_v)$ over all pairs and correcting the first two classes,

$$\begin{aligned} \frac{\mathbb{E}(S_n(S_n-1))}{n(n-1)} = \mathbb{E}[\rho_k(d_u)\rho_k(d_v)] &+ \mathbb{P}(uv \in E)\,\mathbb{E}[\pi_{\text{dist}}(a,0,c) - \rho_k(a+1)\rho_k(c+1) \mid uv \in E] \\ &+ \mathbb{P}(\text{dist}(u,v) = 2)\,\mathbb{E}[(\pi_{\text{same}} + \pi_{\text{dist}})(a,1,c) \\ &\quad -\rho_k(a+1)\rho_k(c+1) \mid \text{dist}(u,v) = 2], \end{aligned} \tag{40}$$

Every term is a bracket $\langle s^a t^c g(s,t)\rangle$ with $g$ independent of $(a, c)$, so the expectation over the tree can be taken inside the bracket, where Lemma 3 gives $\mathbb{E}[s^a t^c]$ in each class: this is Eq. (39). For the asymptotics write $\mathbb{E}(S_n) = n\mu_n$ and $\mathbb{E}(S_n(S_n-1)) = n(n-1)\nu_n$. Uniformly for bounded $z$, $(1 + z/n)^{n-j} = e^z(1 - (z^2/2 + jz)/n + O(n^{-2}))$, so $\mu_n = \mu + \mu_1/n + O(n^{-2})$ and $\nu_n = \nu + \nu_1/n + O(n^{-2})$ with $\mu = k\int_{1-1/k}^1 se^{s-1}\,\mathrm{d}s = e^{-1/k}$, $\nu = \langle ste^{s+t-2}\rangle = \mu^2$,

$$\mu_1 = -k\int_{1-1/k}^{1} se^{s-1}\left(\frac{(s-1)^2}{2}+2(s-1)\right)\mathrm{d}s,$$

$$\nu_1 = -\langle ste^{s+t-2}\left(\frac{(s+t-2)^2}{2}+2(s+t-2)\right)\rangle + \langle (s+t)e^{s+t-2}\left(\frac{k-1}{k}-st\right)\rangle \quad (41)$$
$$+\langle (s+t+1)e^{s+t-2}\left(\frac{\min(s,t)}{k}+\frac{k-1}{k}(s+t-1)-st\right)\rangle.$$

Then $\mathrm{Var}(S_n) = n^2(\nu-\mu^2)+n(\nu_1-\nu+\mu-2\mu\mu_1)+O(1)$, the first term vanishes, and the elementary integrals give $\mu_1 = (2k-1)e^{-1/k}/(2k^2)$ and $\nu_1 = 2(k-1)e^{-2/k}/k^2$, whence $\sigma_k^2 = \nu_1-\mu^2+\mu-2\mu\mu_1 = \mu(1-\mu)-\mu^2/k^2 = e^{-1/k}-(1+k^{-2})e^{-2/k}$. ◀

The value of $\mu_1$ sharpens Eq. (17) to $\mathbb{E}(T_n) = \big(1-e^{-1/k}\big)n-(2k-1)e^{-1/k}/(2k^2)+O(1/n)$, whose constant is $-1/(2e)$ at $k=1$ as in Eq. (16). In the slope, $\mu = e^{-1/k}$ and $\mu/k$ are the probabilities that a Poisson variable of mean $1/k$ equals 0 and 1, so $\sigma_k^2 = p_0(1-p_0)-p_1^2$. Numerically $\sigma_2^2 \approx 0.147$ and $\sigma_3^2 \approx 0.146$, and $\sigma_k^2 = 1/k-5/(2k^2)+O(k^{-3})$, so that the variance and the mean of $T_n$ become equivalent as $k$ grows.

**Proof of Theorem 9.** By Proposition 5 the peak count $S_n$ has the same law as the number of leaves of a uniform random labeled tree. The number of leaves, and more generally the number of vertices of any fixed degree, of a uniform random labeled tree (equivalently a Galton-Watson tree conditioned on its size) obeys a central limit theorem [Drm09]. Its variance is of order $n$ by Corollary 6, so the limit is nondegenerate. Since $T_n = n-S_n$ is affine in $S_n$, it is asymptotically normal with the stated centering and scaling. ◀

▶ **Proposition 11** (Poisson Galton-Watson tree). *For the Galton-Watson tree with* Poisson(1) *offspring, the probability generating function of the number of peaks is*

$$C\big(e^{(v-1)/e-1}\big)+\frac{v-1}{e}, \quad \text{valid for } |v|\le 1. \quad (42)$$

**Proof.** Write $F(z,v) = C\big(ze^{(v-1)z}\big)+(v-1)z$ for Eq. (14), so that $n![z^n]F(z,v) = \sum_{|\tau|=n} v^{\mathrm{peaks}(\tau)}$ over the $n^{n-1}$ rooted labeled trees of size $n$. A Poisson(1) Galton-Watson tree has total-progeny law $\mathbb{P}(|\mathcal{T}|=m) = e^{-m}m^{m-1}/m!$ and, conditioned on its size $m$ and given a uniform random vertex labeling, is a uniform rooted labeled tree (a Cayley tree) [Drm09]. As a peak depends only on the relative order of labels, its peak probability generating function is therefore

$$\sum_{m\ge 1}\mathbb{P}(|\mathcal{T}|=m)\frac{1}{m^{m-1}}\sum_{|\tau|=m} v^{\mathrm{peaks}(\tau)} = \sum_{m\ge 1} e^{-m}[z^m]F(z,v) = F(1/e,v). \quad (43)$$

Since $(1/e)e^{(v-1)/e} = e^{(v-1)/e-1}$, this equals $C\big(e^{(v-1)/e-1}\big)+(v-1)/e$. Equivalently, it is Eq. (14) composed with the Poisson size generating function $A(z) = C(z/e)$. It converges for $|v|\le 1$, where the argument has modulus at most $1/e$, the radius of convergence of $C$.◀

The Poisson(1) tree is critical, so its size, and with it its peak count, has infinite mean. Conditioning on the size $n$ returns the exact law of Theorem 8 and the finite moments of Corollary 6. By Proposition 5 this is also the leaf probability generating function of the Poisson(1) tree.

**General rooted trees.** Consider rooted labeled *plane* trees (ordered children) carrying a weight $\varphi_c$ on each node of out-degree $c$, with $\Phi(z) = \sum_c \varphi_c z^c$, so that the unmarked generating function obeys the classical $A(z) = z\Phi(A(z))$. For the Cayley weights $\varphi_c = 1/c!$ this is the uniform unordered labeled tree. To mark peaks we use a catalytic variable $x$ that records the *holes* of a node (the slots where a further child could be inserted), so a node of out-degree $c$ with $d$ holes weighs $\binom{c+d}{c}\varphi_{c+d}$ and a full child-and-hole set has generating function $\Phi(z+x)$.

▶ **Proposition 12** (General rooted trees and Galton-Watson trees)**.** *Let $A(z,v,x)$ be the generating function of weighted rooted labeled trees in which $z$ marks the nodes, $v$ the peaks and $x$ the holes. It is characterized by*

$$\partial_z A = (v-1)\Phi(x) + \Phi(A+x) + (\partial_x A)\Phi(A+x), \tag{44}$$

*and the monochrome peak generating function is $A(z,v,0)$. For $\Phi(z)=e^z$ it is Eq. (14). If $\Phi$ is a probability generating function of mean at most $1$, other than $\Phi(z)=z$, then $A(1,v,0)$ is the peak probability generating function of the Galton-Watson tree with offspring law $\Phi$. Under a uniform $k$-coloring, the generating function $B_k(z,v)$ of trees whose root has a prescribed color satisfies*

$$B_k(z,v) = A(z,v,(k-1)B_k(z,v)), \tag{45}$$

*and the generating function of all $k$-colored trees is $kB_k(z,v)$.*

**Proof.** Letting $z$ mark nodes, $v$ peaks and $x$ holes, removing the smallest-labeled node gives Eq. (44): the removed node is either the root (first two terms, according to whether it is a peak, which happens iff it has no children) or an internal node whose removal turns its parent slot into a hole (last term). The quantity of interest is $A(z,v,0)$. For the Cayley weights $\varphi_c = 1/c!$, i.e. $\Phi(z)=e^z$, a hole behaves like a marked child slot at a single node and $A(z,v,x) = F(ze^x, v)$ with $F$ as in Eq. (14), which satisfies the equation and reduces at $x=0$ to Eq. (14). If $\Phi$ is a probability generating function of mean at most 1, other than $\Phi(z)=z$ (so that the tree is almost surely finite), then $A(1,v,0)$ is well defined for $|v|\le 1$ and is the peak probability generating function of the Galton-Watson tree with offspring law $\Phi$. The Poisson(1) case recovers Proposition 11. Unlike the Cayley case, peaks and leaves are not equidistributed in general: for the full binary offspring law the tree on three vertices has peak generating function $(2v+v^2)/3$ and leaf generating function $v^2$.

For the random coloring, consider a $k$-colored general rooted tree with holes and keep only the vertices of some fixed color, say *red*. The connected component containing the root is a monochrome (red) general rooted tree with holes, in which each hole either was already a hole in the original tree, or held the root of a $k$-colored subtree whose root is one of the $k-1$ non-red colors. Since a hole and such an attached non-red subtree play the same structural role at a red vertex, a child slot that does not affect the red vertex's peak status, the two combine additively in the holes variable. Writing $B_k(z,v)$ for the generating function of $k$-colored general rooted trees with a red root and no hole, this gives Eq. (45), and summing over the $k$ root colors gives $kB_k(z,v)$. In the Cayley case $B_k = F\big(ze^{(k-1)B_k}, v\big)$, which Lagrange inversion solves (Theorem 10).

**Proof of Theorem 10.** Orient the edges away from the root (peaks do not depend on it) and condition on the set $M$ of initially conflicting edges. Each non-root vertex has the color of its parent with probability $1/k$, independently of the others, so $\mathbb{P}(M=M_0) = k^{-|M_0|}(1-1/k)^{n-1-|M_0|}$ for every $M_0 \subseteq E(T)$, and a vertex is a peak iff it is a local maximum of the forest $(V,M)$, whose components are monochrome: given $M$ with components $\Phi_1,\dots,\Phi_m$ of sizes $c_1,\dots,c_m$, $v^{S_n} = \prod_i v^{\mathrm{peaks}(\Phi_i)}$. The rooted trees containing a given spanning forest $\Phi$ with these components number $n^{m-1}c_1\cdots c_m$, that is $n$ roots times the unrooted count $n^{m-2}c_1\cdots c_m$ [Moo70, Pit99], which is the weighted Cayley formula $\sum_\tau \prod_i x_i^{\deg_\tau(i)} = x_1\cdots x_m(x_1+\cdots+x_m)^{m-2}$ over the trees $\tau$ on the $m$ contracted components, at $x_i = c_i$. Summing over the trees containing $\Phi$, then over the spanning forests with $m$ components,

$$\mathbb{E}\big(v^{S_n}\big) = \frac{1}{n^{n-1}} \sum_{m=1}^{n} k^{-(n-m)}\Big(1-\frac{1}{k}\Big)^{m-1} n^{m-1} \sum_{\Phi}\prod_{i=1}^{m} c_i v^{\mathrm{peaks}(\Phi_i)}. \tag{46}$$

The factor $c_i$ roots the component $\Phi_i$, and rooted labeled trees by peaks have the exponential generating function $F$, so the inner sum is $n![z^n]F(z,v)^m/m!$. Since $k^{-(n-m)}(1-1/k)^{m-1} = (k-1)^{m-1}k^{-(n-1)}$, the sum over $m$ is $n!k^{-(n-1)}[z^n]\sum_{m\geq 1}((k-1)nF)^m/((k-1)n\,m!)$, which is Eq. (19). At $v=1$, $F(z,1)=C(z)$ and $[z^n]e^{aC(z)} = a(a+n)^{n-1}/n!$ [FS09] give $\mathbb{E}(1)=1$. Alternatively, with $u = ze^{(k-1)B_k}$ the fixed point of Proposition 12 reads $z = ue^{-(k-1)F(u,v)}$, and Lagrange inversion gives $[z^n]B_k = [u^n]e^{(k-1)nF(u,v)}/((k-1)n)$. As $n![z^n]B_k$ counts the colored trees whose root has a prescribed color, $\mathbb{E}\big(v^{S_n}\big) = n![z^n]B_k/(k^{n-1}n^{n-1})$, the same formula. Expanded, the right-hand side of Eq. (19) is a polynomial in $k$, whose value at $k=1$ is $n![z^n]F(z,v)/n^{n-1}$, the monochrome law of Theorem 8. ◀

**A binary-tree observation.** On rooted full binary *plane* trees with $2m+1$ labeled vertices, the number of labelings with exactly one peak appears to be $(2m)![z^{2m}]\cos(z)^{-4}$, the sequence OEIS A395382: we checked the identity for $m \leq 10$ but have not proved it.

## E Erdős-Rényi graphs: proofs and variance

**Proof of Proposition 6.** List the vertices by decreasing label as in Section 5, so that the $j$-th vertex has the $j$-th largest label. It is a peak with probability $\mathbb{E}\big[(1-p)^{h_j}\big]$ with $h_j \sim \text{Binomial}(j-1, 1/k)$, that is $(1-p/k)^{j-1}$. Summing over $j$ gives $\mathbb{E}(S) = (k/p)(1-(1-p/k)^n)$ and $\mathbb{E}(T) = n - \mathbb{E}(S)$. This expression depends on $p$ and $k$ only through $p/k$, so the mean under a uniform $k$-coloring at edge probability $p$ equals the monochrome mean at edge probability $p/k$. Indeed, a vertex of larger label than the $j$-th prevents it from being a peak iff it shares its color and is adjacent to it, an event of probability $p/k$, as in the monochrome case at $p/k$. This equality fails for the variance, which depends on pairs of vertices. The conflicting edges form $k$ separate random graphs, one per color class, not a single $G(n, p/k)$: two vertices of distinct colors never interact. For instance, at $n=20$, $p=0.2$, $k=2$ the variance is 3.38, against 3.60 monochrome at $p=0.1$. ◀

The same reading gives the second factorial moment in a few lines.

▶ **Proposition 13** (Erdős-Rényi, second factorial moment). *On $G(n,p)$ with $k$ colors, with $s = 1-p/k$, $t_1 = 1-(2p-p^2)/k$, $t_2 = 1-2p/k$ and $F(g) = (k/p)[(1-g^n)/(1-g) - (s^n - g^n)/(s-g)]$,*

$$\mathbb{E}(S(S-1)) = 2\left[\frac{1-p}{k}F(t_1) + \frac{k-1}{k}F(t_2)\right]. \tag{47}$$

*At $k=1$ this is $\left(\sum_{i<n}(1-p)^i\right)^2 - \sum_{i<n}(1-p)^{2i}$, the second factorial moment of the independent sum of Proposition 7.*

**Proof.** Given the colors the peak indicators are independent, so $\mathbb{E}(S(S-1)) = 2\sum_{j<j'}\mathbb{E}\Big[(1-p)^{h_j+h_{j'}}\Big]$, the expectation being over the colors. If the $j$-th and $j'$-th vertices share their color (probability $1/k$), each of the $j-1$ vertices of larger label than the $j$-th is counted in both $h_j$ and $h_{j'}$ with probability $1/k$, whence a factor $(k-1)/k + (1-p)^2/k = t_1$ each; each of the $j'-j-1$ vertices in between is counted in $h_{j'}$ with probability $1/k$, whence a factor $s$; and the $j$-th vertex itself is counted in $h_{j'}$, whence the factor $1-p$. If their colors differ (probability $(k-1)/k$), a vertex of larger label than the $j$-th is counted once with probability $2/k$, whence a factor $t_2$, the vertices in between contribute $s$ each and the $j$-th vertex 1. Summing the geometric series over $j<j'$ gives Eq. (47). ◀

**Random graph and fixed graph.** The generating functions Eq. (20) and Eq. (21) and every random-graph formula of the paper describe the joint law over the graph, the coloring and the labeling. For a fixed graph the peak indicators are dependent, and the joint variance is the mean, over the graph, of the variance on a fixed graph, plus the variance, over the graph, of the mean on a fixed graph. Only the first term measures the fluctuations of the stabilization time on one fixed network. The second measures the differences between networks.

**Palette on random graphs.** On $G(n,p)$ the maximum degree $\Delta$ is random, so the condition $q \geq \Delta + 1$ concerns the palette relative to a random quantity. Taking $q = n$ always satisfies it. More economically, for *constant* $p$ one has $\Delta = (1 + o(1))np$ with high probability [JŁR00], so any palette $q \geq (1 + \varepsilon)np$ meets $q \geq \Delta + 1$ with probability $1 - o(1)$. On that event $T = n - S$ holds exactly and the excluded draws contribute a vanishing correction. When $np$ stays bounded, this fails: the mean degree is constant but $\Delta = \Theta(\log n/ \log\log n)$, so the palette must grow at least that fast. The peak statistic $S$, and hence every mean and variance formula, is defined for any $k$ with no palette constraint. The bound $q \geq \Delta + 1$ is only what ties it to the algorithm's recoloring count.

**Variance through the second factorial moment theorem.** Unlike the triangle-free families, $G(n,p)$ has triangles, so its variance exercises the full Theorem 3, and the computation below is the one that the bipartite model reuses (Appendix F). Averaging over the random graph makes the neighborhood profile of a pair explicit: for a random ordered pair $(u, v)$, each of the other $n - 2$ vertices is independently a common neighbor with probability $\beta = p^2$, a private neighbor of $u$ (resp. $v$) with probability $\alpha = p(1 - p)$, or neither with probability $\delta = (1 - p)^2$, so $(a, b, c)$ is multinomial. Since $\pi_{\text{same}}$ and $\pi_{\text{dist}}$ are expectations of products whose exponents are exactly $a, b, c$, this multinomial average collapses the sum of Eq. (5) into closed forms.

▶ **Proposition 14** (Erdős-Rényi, variance). *On $G(n,p)$ with $k$ colors, write $\alpha = p(1-p)$, $\beta = p^2$, $\delta = (1-p)^2$. The variance of the stabilization time is* $\mathrm{Var}(T) = \mathbb{E}(S(S-1)) + \mathbb{E}(S) - \mathbb{E}(S)^2$ *with* $\mathbb{E}(S) = (k/p)(1 - (1 - p/k)^n)$ *as in Eq. (22) and*

$$\mathbb{E}(S(S-1)) = n(n-1)[I_{\text{dist}} + (1-p)I_{\text{same}}], \tag{48}$$

*where, with $m = n - 2$, $I_{\text{dist}}$ and $I_{\text{same}}$ are the closed forms*

$$\begin{aligned} I_{\text{dist}} &= \frac{k-1}{k}\frac{(A+2B)^{m+2} - 2(A+B)^{m+2} + A^{m+2}}{B^2(m+1)(m+2)}, \\ I_{\text{same}} &= \frac{2}{(\alpha+\beta)(m+1)(m+2)}\left(\frac{1 - C^{m+2}}{E} - \frac{(C+D)^{m+2} - C^{m+2}}{D}\right), \end{aligned} \tag{49}$$

*with $A = \delta + (2\alpha(k-1) + \beta(k-2))/k,\ B = (\alpha+\beta)/k,\ C = (k-1)(2\alpha+\beta)/k + \delta,\ D = \alpha/k$, and $E = (2\alpha+\beta)/k$ (so $C + E = 1$). This is the* joint *variance over the random graph, the random coloring, and the random labeling.*

**Proof.** Adjacency of $u$ and $v$ is independent of the other vertices, so a pair is adjacent with probability $p$ (contributing $\pi_{\text{dist}}$ only, by Theorem 3) and non-adjacent with probability $1 - p$ (contributing $\pi_{\text{same}} + \pi_{\text{dist}}$), hence $\mathbb{E}(S(S-1)) = n(n-1)\ \mathbb{E}[\pi_{\text{dist}} + (1-p)\pi_{\text{same}}]$. Taking the multinomial expectation of Eq. (26) and using $\mathbb{E}\left[\xi_a^a \xi_b^b \xi_c^c\right] = \left(\alpha\xi_a + \beta\xi_b + \alpha\xi_c + \delta\right)^{n-2}$ leaves an integrand affine in each variable, which $\int_0^1 (A + Bt)^m\,\mathrm{d}t = ((A+B)^{m+1} - A^{m+1})/(B(m+1))$, applied once for $I_{\text{dist}}$ and twice for $I_{\text{same}}$, evaluates to the closed forms above (the two symmetric $\pi_{\text{same}}$ terms coincide because $\alpha$ multiplies both private counts). In integral form,

$$I_{\text{dist}} = \frac{k-1}{k}\int_0^1\int_0^1\left(\delta + \frac{2\alpha(k-1)+\beta(k-2)+(\alpha+\beta)(x+y)}{k}\right)^m \mathrm{d}x\,\mathrm{d}y,$$
$$I_{\text{same}} = \frac{2}{k}\int_0^1\int_0^1 y\left((\alpha+\beta)\frac{k-1+xy}{k} + \alpha\frac{k-1+y}{k} + \delta\right)^m \mathrm{d}x\,\mathrm{d}y. \quad (50)$$

◀

**Proof of Proposition 7.** The factorization of Eq. (20) is the probability generating function of the independent sum $\sum_i B_i$. For fixed $p$, $\sum_{i\geq 0}(1-p)^i < \infty$, so by Borel-Cantelli only finitely many $B_i$ equal 1 almost surely and $S_n \to S_\infty$, whose mean is $\sum_i (1-p)^i = 1/p$. When $p_n \to 0$ with $np_n \to \infty$ the summands are bounded and $\mathrm{Var}(S_n) = \sum_{i<n}(1-p)^i(1-(1-p)^i) \sim 1/(2p)$ diverges, so Lindeberg's condition holds and $S_n$ is asymptotically normal. With $k$ colors, condition on the color classes: they induce vertex-disjoint subgraphs with independent edge sets, a class of size $m$ carries a peak count distributed as Eq. (20) with $n$ replaced by $m$, and all class sizes tend to infinity almost surely (with $mp \to \infty$ in the second regime), so the $k$ conditionally independent counts converge jointly to $k$ independent copies of the monochrome limit. In the second regime the fluctuation of the conditional mean must be compared with the conditional variance, which is $(1+o(1))k/(2p)$ in probability, while, by the Berry-Esseen bound for independent bounded indicators, the conditional law is within Kolmogorov distance $O\big(\sqrt{p/k}\big)$ of the normal law, uniformly over the class sizes. Writing $N_i$ for the class sizes, the conditional mean is $\sum_i (1-(1-p)^{N_i})/p$, so its deviation from $\mathbb{E}(S_n)$ is $D = -(1/p)\sum_i[(1-p)^{N_i} - (1-p/k)^n]$ and, by Cauchy-Schwarz, $\mathrm{Var}(D) \leq (k^2/p^2)\,\mathrm{Var}((1-p)^N)$ with $N \sim \mathrm{Binomial}(n, 1/k)$. Now $\mathrm{Var}((1-p)^N) = (1-(2p-p^2)/k)^n - (1-p/k)^{2n}$, a difference of $n$-th powers of two numbers at distance $p^2(k-1)/k^2$, hence at most $np^2(k-1)k^{-2}e^{-2(n-1)p(1-p/2)/k} \leq np^2(k-1)k^{-2}e^{-np/k}$ for $p \leq 1/2$ and $n \geq 3$. Therefore $\mathrm{Var}(D)/\mathrm{Var}(S_n) = O\big((k-1)np\,e^{-np/k}\big)$, which tends to 0 when $np \to \infty$, and the conditionally Gaussian limit is Gaussian. ◀

At fixed $p$ the mean degree grows with $n$, so the expected number of peaks $\mathbb{E}(S) \to k/p$ stays bounded and the variance tends to a constant rather than growing linearly (visible in Figure 2). The closed forms track this limit exactly. The companion code checks these formulas.

## F Bipartite graphs: proofs and variance

**Proof of Proposition 8.** As in Eq. (22), with $\mathbb{E}[t^d] = (1-p+pt)^{n_2}$ on the first side and $(1-p+pt)^{n_1}$ on the second, weighted by $n_1$ and $n_2$: integrate $k\int_{1-\frac{1}{k}}^1$ term by term. ◀

**Variance.** Same-side pairs (both in the first side, or both in the second) are non-adjacent and have their common and private neighbors among the opposite side, so they reuse the $G(n,p)$ integrals $I_{\text{dist}}(m)$ and $I_{\text{same}}(m)$ of Eq. (48) with the exponent $n-2$ replaced by the opposite-side size. Cross pairs (one endpoint on each side) can have no common neighbor, so $b = 0$. Their private-neighbor counts are independent binomials, and adjacency occurs with probability $p$.

▶ **Proposition 15** (Bipartite Erdős-Rényi, variance). *Write $I_{\text{dist}}(m)$, $I_{\text{same}}(m)$ for the closed forms of Eq. (48), set $r = 1 - p/k$, and*

$$\begin{aligned}
L_{\text{dist}} &= (k-1)\frac{k}{p^2 n_1 n_2}(1-r^{n_1})(1-r^{n_2}), \\
L_{\text{same}} &= \Lambda(n_2-1, n_1-1) + \Lambda(n_1-1, n_2-1), \text{with} \\
\Lambda(M, M') &= \frac{k}{p^2(M+1)}\left(\frac{1-r^{M+M'+2}}{M+M'+2} - r^{M+1}\frac{1-r^{M'+1}}{M'+1}\right).
\end{aligned} \tag{51}$$

*Then, with* $\mathbb{E}(S)$ *as in Eq. (23),*

$$\begin{aligned}
\mathbb{E}(S(S-1)) = n_1(n_1-1)&[I_{\text{same}}(n_2) + I_{\text{dist}}(n_2)] \\
&+n_2(n_2-1)[I_{\text{same}}(n_1) + I_{\text{dist}}(n_1)] \\
&+2n_1 n_2[pL_{\text{dist}} + (1-p)(L_{\text{same}} + L_{\text{dist}})],
\end{aligned} \tag{52}$$

*and* $\operatorname{Var}(T) = \mathbb{E}(S(S-1)) + \mathbb{E}(S) - \mathbb{E}(S)^2$. *In integral form,*

$$\begin{aligned}
L_{\text{dist}} &= \frac{k-1}{k}\int_0^1\int_0^1 \left(1 - \frac{p(1-x)}{k}\right)^{n_2-1}\left(1 - \frac{p(1-y)}{k}\right)^{n_1-1} \mathrm{d}x\,\mathrm{d}y, \\
L_{\text{same}} &= \frac{1}{k}\int_0^1\int_0^1 y\left[\left(1 - \frac{p(1-xy)}{k}\right)^{n_2-1}\left(1 - \frac{p(1-y)}{k}\right)^{n_1-1}\right. \\
&\qquad \left. + \left(1 - \frac{p(1-xy)}{k}\right)^{n_1-1}\left(1 - \frac{p(1-y)}{k}\right)^{n_2-1}\right] \mathrm{d}x\,\mathrm{d}y.
\end{aligned} \tag{53}$$

The companion code checks this formula.

**Proof of Proposition 9.** *Conditional independence.* Attribute each edge to its endpoint of smaller label. Ordering the vertices by decreasing label, the $j$-th vertex is a peak iff none of its edges to the $h_j$ opposite-side vertices of larger label that share its color is present. These edge sets are disjoint for distinct $j$, so conditionally on the sequence of sides and colors the peak indicators are independent with the stated probabilities, which gives the stated generating function.

*Limit at fixed p, monochrome.* Write $S_n = \sum_{j\geq 1} B_j$ with $B_j$ the indicator that the vertex of $j$-th largest label is a peak. Given the sides, $h_j$ counts the vertices of the opposite side among the $j-1$ vertices of larger label. The sequence of sides is a uniform arrangement of $n_1$ first-side and $n_2$ second-side symbols, so for a first-side vertex $h_j$ is hypergeometric (a sample of size $j-1$ without replacement from the other $n-1$ vertices, $n_2$ of which lie on the second side), with mean $(j-1)n_2/(n-1)$, and symmetrically. By Hoeffding's comparison of sampling without and with replacement for convex functions [Hoe63], $\mathbb{E}\big[(1-p)^{h_j}\big] \leq (1-pn_2/(n-1))^{j-1}$ for a first-side vertex and $\mathbb{E}\big[(1-p)^{h_j}\big] \leq (1-pn_1/(n-1))^{j-1}$ for a second-side one. Both are bounded by $\theta^{j-1}$ with $\theta = 1 - p\min(\lambda, 1-\lambda)/2 < 1$ for all $n \geq n_0$ and all $j$, since $n_1/n \to \lambda \in (0,1)$. Hence $\sum_{j>J}\mathbb{E}(B_j) \leq \theta^J/(1-\theta)$ uniformly in $n$: the tail of $S_n$ beyond the top $J$ vertices is negligible in probability, uniformly in $n$. For fixed $J$ the top $J$ sides converge in law to i.i.d. Bernoulli($\lambda$) draws, so $\sum_{j\leq J} B_j$ converges in law to its i.i.d. counterpart. Letting $J \to \infty$ gives $S_n \to S'_\infty = \sum_{j\geq 1} B_j^\infty$, where, given i.i.d. sides with $\mathbb{P}(\text{first side}) = \lambda$, the $B_j^\infty$ are independent Bernoulli$\big((1-p)^{h_j}\big)$. The limit $S'_\infty$ is almost surely finite since its mean is. The mean follows: a first-side vertex at position $j$ has $h_j \sim \text{Binomial}(j-1, 1-\lambda)$, so $\mathbb{E}\big[(1-p)^{h_j}\big] = (1-p(1-\lambda))^{j-1}$, and summing over $j$ with weight $\lambda$, plus the symmetric term, gives $\lambda/(p(1-\lambda)) + (1-\lambda)/(p\lambda)$.

*k colors.* Conditionally on the colors, the color classes induce vertex-disjoint bipartite subgraphs with independent edge sets, each with sides of sizes $n_1/k + O(\sqrt{n})$ and $n_2/k +$ $O(\sqrt{n})$ in probability, hence with the same side proportion $\lambda$ in the limit. The $k$ conditionally independent peak counts converge jointly to $k$ independent copies of $S'_\infty$. ◀

## G Numerical checks

Two independent programs were used. The first runs the actual recoloring dynamics under the random central daemon (an event loop that draws a uniformly random enabled process at each step and recolors it with a random free color) and records the number of moves. It does not use Lemma 1 and serves to check it. The second is a *peak-oracle* sampler: by Lemma 1 one sample of the stabilization time needs only a coloring and a labeling, not a run of the dynamics, which makes large campaigns cheap. On fourteen test graphs (at most seven vertices, listed with the companion code), the law of the move count under the first agrees with the exact law of $n - S$, computed by enumeration, for $k = 1, 2, 3$ initial colors. In addition, every generating function and closed form of the paper was compared with exact enumeration for small sizes:

- the mean and the second factorial moment (Theorem 1, Theorem 3) against enumeration on the test graphs at $k = 1, 2, 3$,
- the path series Eq. (11) and the cycle series for $n \leq 8$, and their $k$-colored versions (Theorem 7) for $n \leq 7$ at $k = 2$ and $n \leq 6$ at $k = 3$,
- the $G(n, p)$ mean and second factorial moment (Proposition 13, hence Proposition 14) against enumeration over all graphs on at most five vertices at $k \leq 3$,
- the catalytic equation Eq. (44), solved as a series: against Eq. (14) for the Cayley weights, and against the enumeration of rooted trees on $n \leq 5$ vertices for three other weight systems, monochrome and through the fixed point Eq. (45) for $k \leq 3$,
- the binary-tree count for $m \leq 10$, by enumeration of the labelings for $m \leq 4$ and beyond with the hook-length formula (a labeling has a single peak iff its labels decrease away from its maximum),
- the law of the star (Proposition 4) for $n \leq 7$,
- the exact-maximizer statement (Theorem 6) and Theorem 4 by exhaustive enumeration of all graphs on $n \leq 7$ vertices, the bound Eq. (29) on those without isolated vertices and on random and structured graphs up to $n = 30$, and, as a consistency check, a simulated-annealing search for $8 \leq n \leq 71$, which finds no graph above the star.

The following were also checked:

- the $k$-colored tree mean Eq. (17) against a Monte Carlo over Prüfer sequences ($(n, k) = (8, 2), (12, 3), (30, 1), (30, 2)$),
- the pair laws of Lemma 3 against the exhaustive enumeration of labeled trees for $n \leq 7$,
- the $k$-colored tree variance (Proposition 10) against exact enumeration over trees, labelings and colorings ($n \leq 6$) and a Prüfer Monte Carlo ($n = 30, 100$),
- the $k$-colored tree law Eq. (19) against the same enumeration ($n \leq 7$, $k = 2, 3$),
- the limit laws of Proposition 7 and Proposition 9 against the empirical laws at large $n$, and, monochrome, the exact law of $S_n$ in Proposition 7 against both regimes,
- the statements of Proposition 2 against the actual dynamics from fixed configurations and against the two ends of the range of the move count, attained by explicit labelings, on random graphs with twelve vertices.

The figures below plot the closed forms against the peak-oracle Monte Carlo across $n$ for all five families, monochrome (Figure 1, Figure 2) and under a random 3-coloring (Figure 3, Figure 4). Each point averages $6 \times 10^4$ samples and agreement is within sampling error throughout. Figure 5 compares the exact law of the monochrome path with its Gaussian limit at $n = 60$. The simulation code, the closed-form implementations and the code of every check above and of the figures are in the companion notebook of the Python package tinto, version 0.1.0, `docs/companion/exact_stabilization_time.ipynb`, archived by Software Heritage as `swh:1:rev:79d72a3026ff486616ef0ba83b70e192de339cee`. Its current version, executed, is in the documentation of tinto, balouf.github.io/tinto.

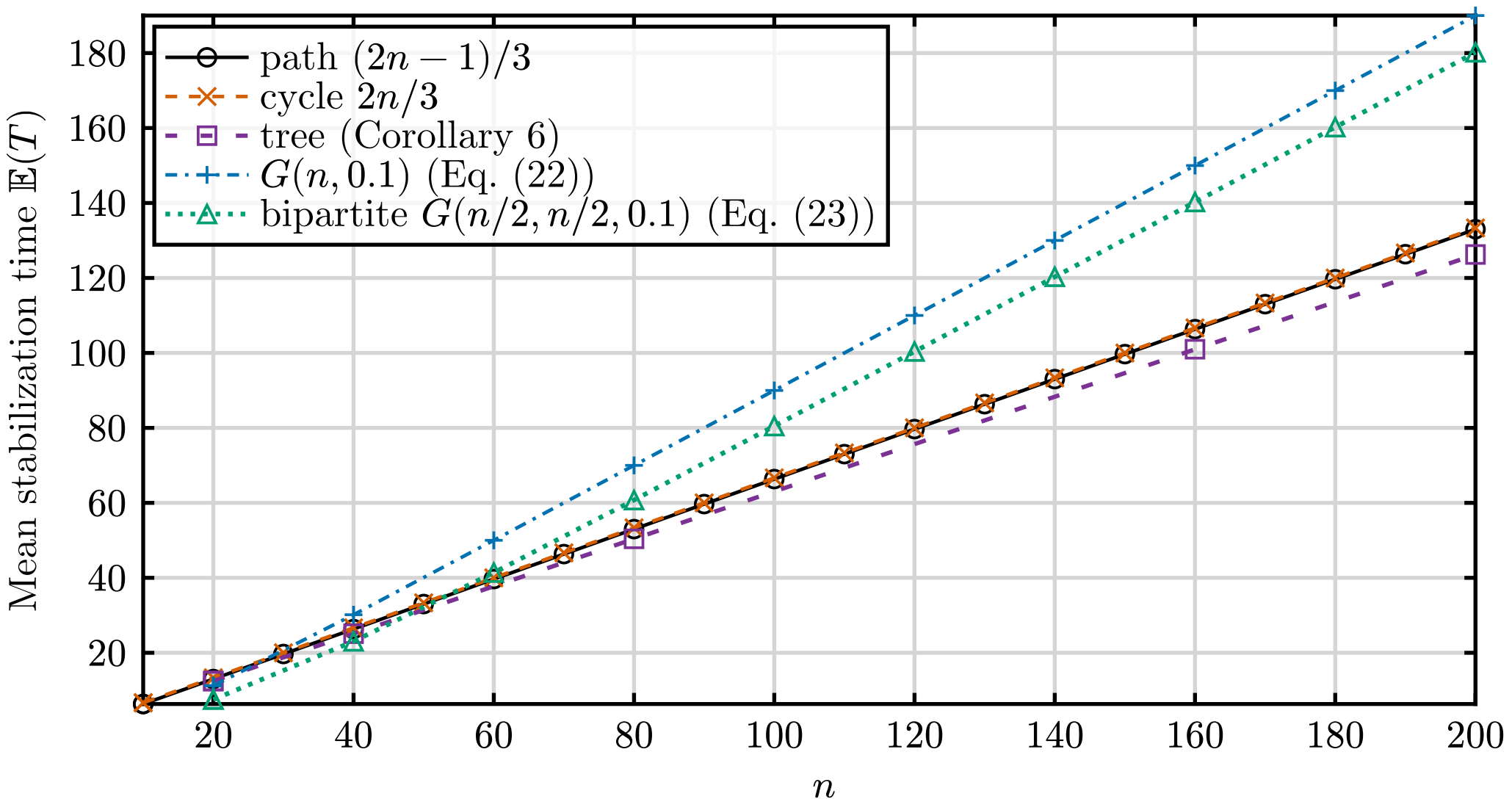

**Figure 1** Mean stabilization time: closed forms (lines) against Monte Carlo (markers), monochrome initial coloring, for the path, cycle, random tree, $G(n, 0.1)$ and balanced bipartite $G(n/2, n/2, 0.1)$. The path and cycle lines coincide, the two means differing by 1/3.

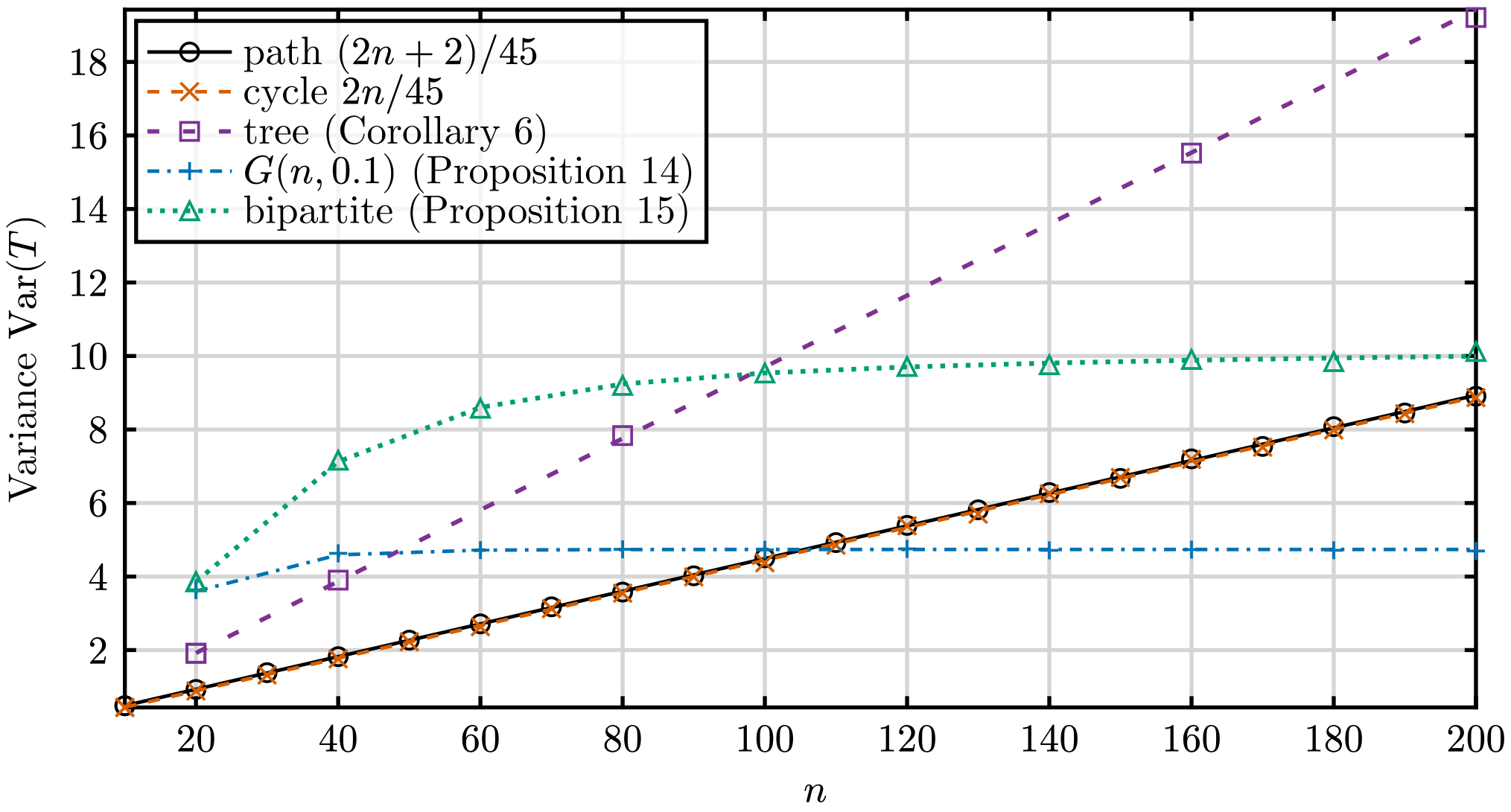

**Figure 2** Variance of the stabilization time: closed forms (lines) against Monte Carlo (markers), monochrome initial coloring. For the bounded-degree families (path, cycle, tree) the variance grows linearly in $n$. For the fixed-density random-graph families $G(n, 0.1)$ and bipartite $G(n/2, n/2, 0.1)$, whose mean degree grows so that the peak count stays bounded, it tends to a constant.

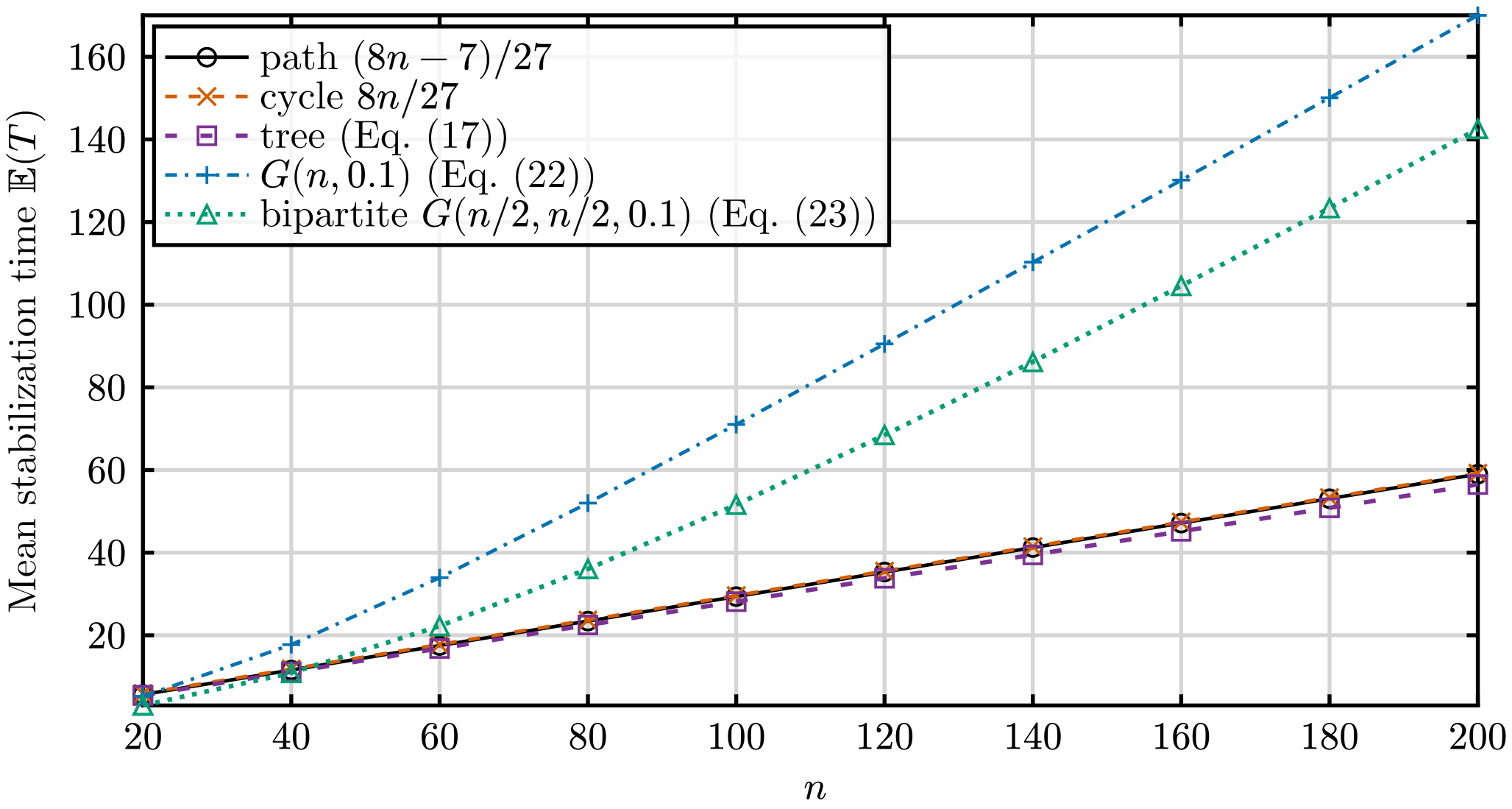

**Figure 3** Mean stabilization time under a random 3-coloring: closed forms (lines) against Monte Carlo (markers), for the path, cycle, random tree, $G(n, 0.1)$ and balanced bipartite $G(n/2, n/2, 0.1)$.

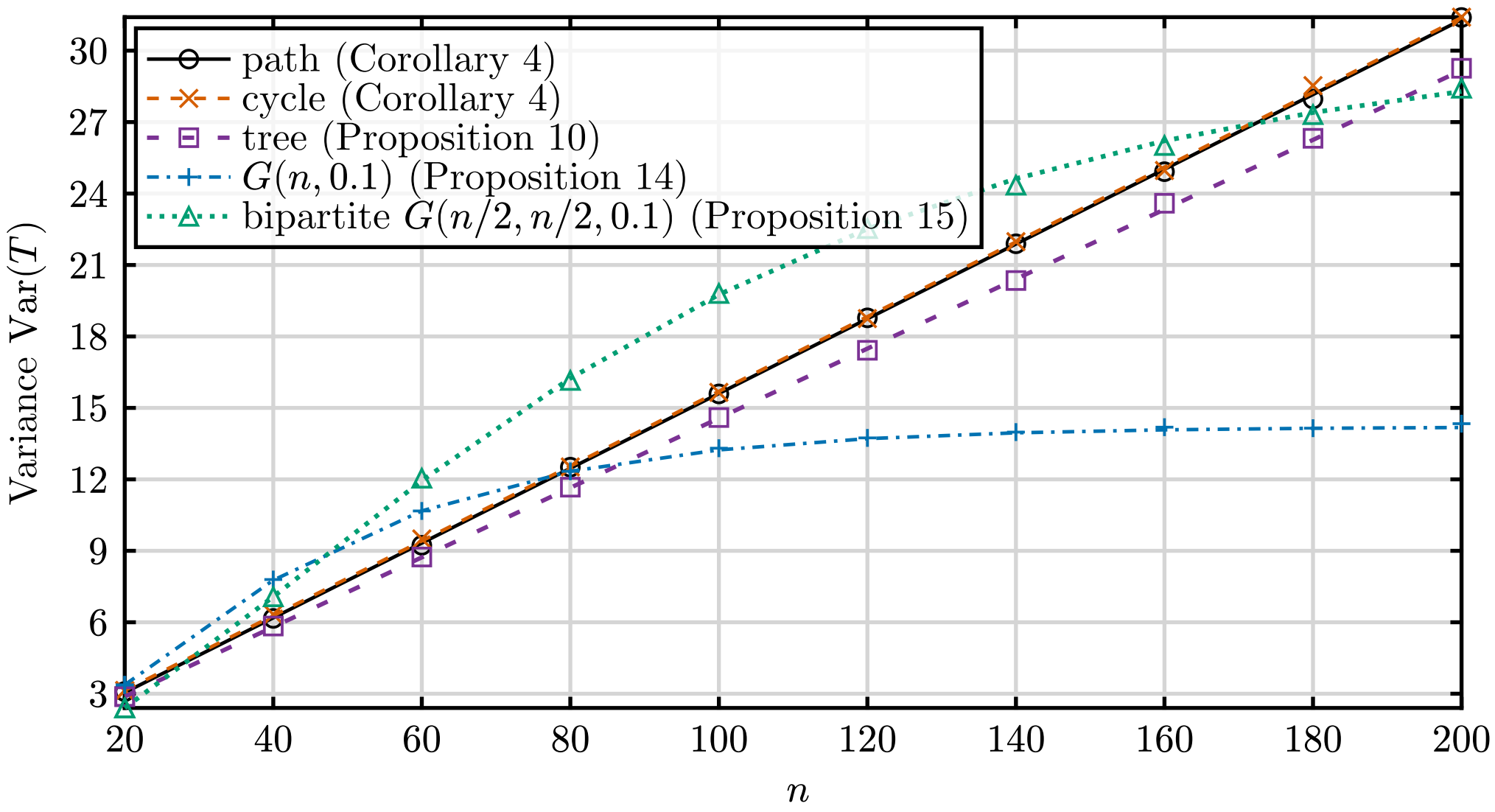

**Figure 4** Variance under a random 3-coloring: closed forms (lines) against Monte Carlo (markers), for the path, cycle, random tree (exact double integral of Proposition 10, with slope $\sigma_3^2 \approx 0.146$ by Eq. (18)), $G(n, 0.1)$ and balanced bipartite $G(n/2, n/2, 0.1)$.

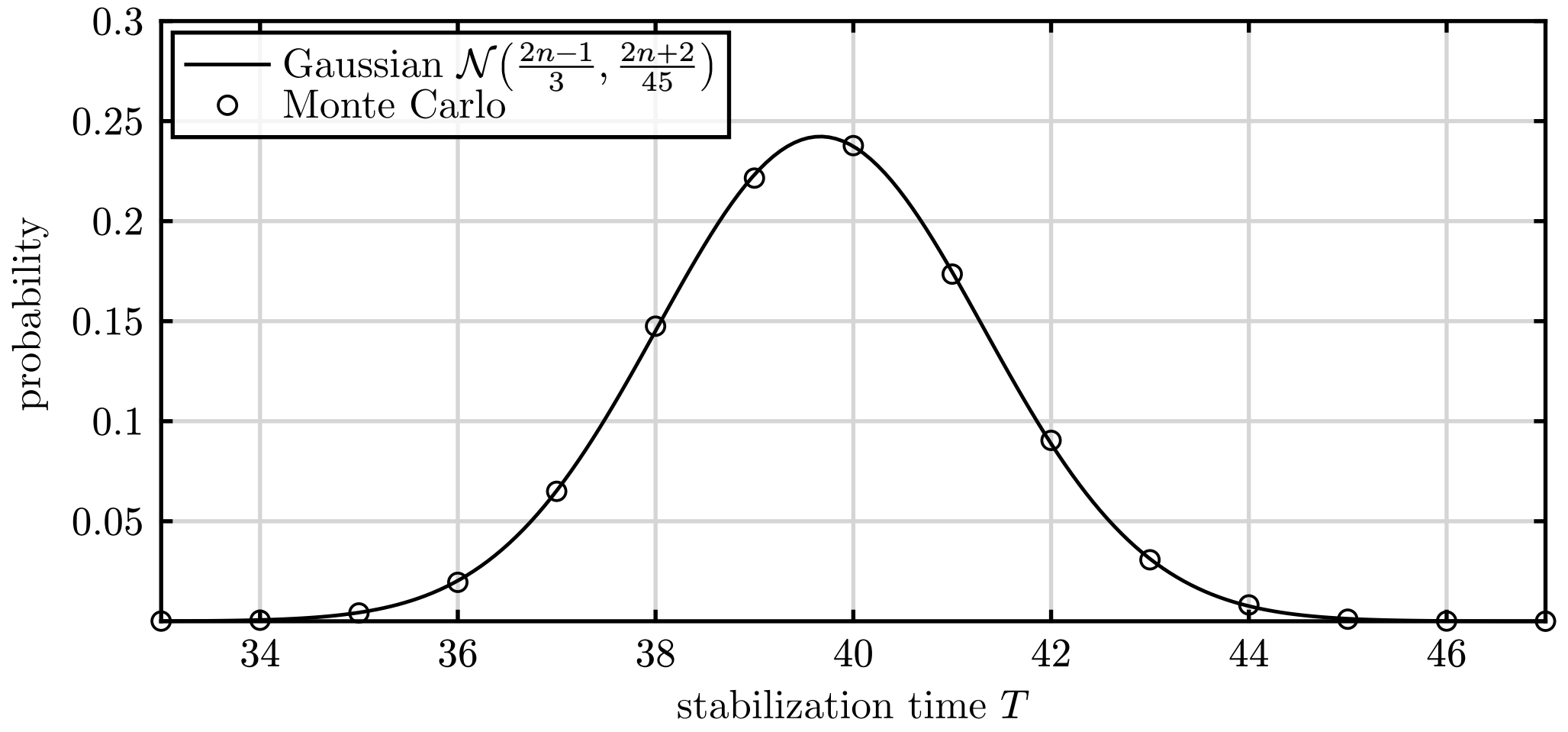

**Figure 5** Empirical law of the stabilization time on the monochrome path at $n = 60$ (Monte Carlo, markers) against the Gaussian limit of Corollary 5 (mean $(2n-1)/3 = 39.67$, variance $(2n+2)/45 \approx 2.71$).